\documentclass{pasj01}
\usepackage{mathrsfs} 
\usepackage{graphicx} 
\usepackage{mediabb}
\usepackage{color}  
  
\title{Spin Temperature and Density of Cold and Warm HI in the Galactic Disk -- Hidden HI --}
 
\author{Yoshiaki \textsc{Sofue}\altaffilmark{1}  }
\altaffiltext{1}{Insitute of Astronomy, The University of Tokyo, Mitaka, Tokyo 181-0015, Japan } 
\email{sofue@ioa.s.u-tokyo.ac.jp}


\begin{document} 
\date{ } 
\maketitle   

\def\vlsr{V_{\rm LSR}} \def\deg{^\circ} \def\r{\bibitem[]{}}     
\def\kms{km s$^{-1}$}  \def\Vsun{V_0}  \def\Vrot{V_{\rm rot}}  
\def\vr{v_{\rm r}} \def\exp{{\rm exp}} 
\def\Xhi{X_{\rm HI}} \def\nH{n_{\rm H}}   
\def\Hcc{ H cm$^{-3}$ } \def\Ihi{I_{\rm HI}}
\def\Tg{ T_{\rm G} } \def\Ts{T_{\rm S}} \def\Tcmb{T_{\rm CMB}} 
\def\Tc{T_{\rm C}} \def\Tb{T_{\rm B}} 
\def\sin{{\rm sin}\ } \def\cos{{\rm cos}\ } 
\def\tan{{\rm tan}\ } \def\cot{{\rm cot}\ } 
\def\red{\textcolor{red}}  \def\Mc{ McClure-Griffiths et al. }
\def\Hcc{ H cm$^{-3}$ } \def\Hsqcm{ H cm$^{-2}$ }
\def\vt{ v_{\rm t} } \def\exp{{\rm exp}\ } \def\X{X_{\rm HI}} 
\def\/{\over} \def\sigv{{\sigma_v}} \def\nsig{n_{\sigma_v}}
\def\Rmax{R_{\rm max}} \def\Rmin{R_{\rm min}} \def\Rwidth{R_{\rm width}}
\def\dlnv{{d\ {\rm ln}\ V\/d\ {\rm ln}\ R}}
\def\D{|R-R_{\rm arm}|} \def\mh{m_{\rm H}} \def\Msun{M_\odot}

\begin{abstract}
We present a method to determine the spin temperature $\Ts$ and volume density $n$ of HI gas simultaneously along the tangent-point circle of galactic rotation in the Milky Way by using the $\chi^2$ method. The best-fit $\Ts$ is shown to range either in $\Ts \sim 100-120$ K or in $1000-3000$ K, indicating that the gas is in cold HI phase with high density and large optical depth, or in warm HI with low density and small optical depth. Averaged values at $3\le R \le 8$ kpc are obtained to be $\Ts=106.7 \pm 16.0$ K and $n=1.53\pm 0.86$ \Hcc for cold HI, and $1720 \pm 1060$ K and $0.38 \pm 0.10$ \Hcc for warm HI, where $R=8\ |\sin \ l|$ kpc is the galacto-centric distance along the tangent-point circle. The cold HI appears in spiral arms and rings, whereas warm HI in the inter-arm regions. The cold HI is denser by a factor of $\sim 4$ than warm HI. The present analysis has revealed the hidden HI mass in cold and optically thick phase in the galactic disk. The total HI mass inside the solar circle is shown to be greater by a factor of $2 - 2.5$ than the current estimation by optically thin assumption. 
\end{abstract}   

\section{Introduction}

The HI spin temperature $\Ts$ in the Milky Way has been determined by emission and absorption observations of the 21-cm line from interstellar clouds in front of radio continuum sources such as HII regions, supernova remnants and extragalactic radio sources
 (Dickey et al. (1978, 2003);
Mebold et al. 1982; 
Kuchar and Bania 1990;  
Liszt 1983, 2001; Liszt et al. 1993;
Roberts et al.1993; 
Stark et al. 1994; 
Wolfire et al. 1995;  
Heiles and Troland 2003a, b; 
Li and Goldsmith 2003; 
Goldsmith and Li 2005; 
Roy et al. 2013a,b;
Brown et al. 2014; Chengalur et al. 2013;
Fukui et al. 2014, 2015;
Murray et al. 2015; 
). 
In these studies, the observed directions are restricted to individual continuum sources, so that the widely distributed HI gas in the galactic disk has been not thoroughly investigated.

In our recent paper (Sofue 2017a) we developed a method to determine the spin temperature and density of diffusely distributed HI gas simultaneously in the velocity-degenerate (VDR) region using the least squares method (hereafter least $\chi^2$ method). Applying the VDR method to local HI at $\vlsr\sim 0$ \kms, we obtained $\Ts \sim 145$ K and HI density $n=0.89$ \Hcc. The method made it possible to determine the global variation of $\Ts$ of diffusely distributed HI.  

In this paper we extend the VDR-least $\chi^2$ method to a more general case toward tangential directions of the Galactic rotation at terminal velocities inside the solar circle. We apply the method to determine the spin temperature, $\Ts$, and volume density, $n$, at various galacto-centric distances, $R$. 

We use the result to discuss the ISM physics of the cold and warm HI gases in the galactic disk on the phase diagrams for interstellar thermal equilibrium (Field et al. 1969; Shaw et al. 2017; Wolfire et al. 2003). We further discuss the distributions of the two phased gases with the the spiral arms and galactic rings (e.g. Burton 1976; Marasco et al. 2017). 

In this paper we call the HI gas with $\Ts\sim 100$ K cold HI, and that with $>\sim 1000$ K warm HI. The spin tempereratue here is that appearing in the radiative transfer equation through the absorption coefficient. Therefore, it is not directly related to the kinetic temperature coupled with the dynamical pressure of cold neutral matter (WNM) and/or warm neutral matter (CNM) (literature as above). However, when the gas is in thermal equilibrium, the cold HI may represent the CNM and warm HI the WNM. 

For analysis we make use of the Leiden-Argentine-Bonn (LAB) all-sky HI survey (Kalberla et al. 2005) and the Parkes Galactic HI survey (GASS) (\Mc 2009; Kalberla et al. 2010) for the HI brightness temperatures, and the 1420 MHz Stockert-Villa-Elisa (SVE) all sky survey (Reich et al. 2001) for the radio continuum brightness near the HI line frequency. We adopt the solar constants of $R_0=8$ kpc and $V_0=238$ \kms (Honma et al. 2012). 

\section{Tangent-Point Method to measure the Spin Temperature and Density} 
\label{sectionTPM}
 
 \subsection{Basic equations}
 
Given the spin temperature ${\Ts}$ and hydrogen density $n$, 
the observed HI brightness temperature $\Tb(v)$ at radial velocity $v$ is given  by
\begin{equation} 
\Tb(v)= ({\Ts}-\Tc) (1-e^{-\tau(v)}).
\label{eqTb} 
\end{equation} 
Here, $\tau$ is the optical depth related to the absorption coefficient $\kappa$ of the HI line in the frequency range $\Delta \nu$ transmitting through a distance $\Delta r$ of a cloud or a region under consideration:
\begin{equation}
\tau(v) =\kappa  \Delta r,
\end{equation}
where the absorption coefficient is given by
\begin{equation}
\kappa=2.601\times 10^{-15} {n \/ \Ts} {1\/ \Delta \nu}\ [{\rm cm^{-1}}]
\label{eqkappa}
\end{equation}
in cgs units and $\Ts$ in K (e.g. van de Hulst et al 1954). 
Rewriting $\kappa$ by velocity as $\Delta \nu=1.42\times 10^9 [{\rm Hz}] \Delta v/c$, we obtain
\begin{equation}
\tau={n  \/ \Xhi {\Ts}}{\Delta r \/ \Delta v}.
\label{eqtau}
\end{equation} 
Here, $\Xhi=1.822\times 10^{18}$ H cm$^{-2} [{\rm km\ s^{-1}}]^{-1}$ is the conversion factor from the optical depth to density ($\Ts$ in K, $v$ in \kms and $r$ in cm).

\begin{figure} 
\begin{center}     
\includegraphics[width=8cm]{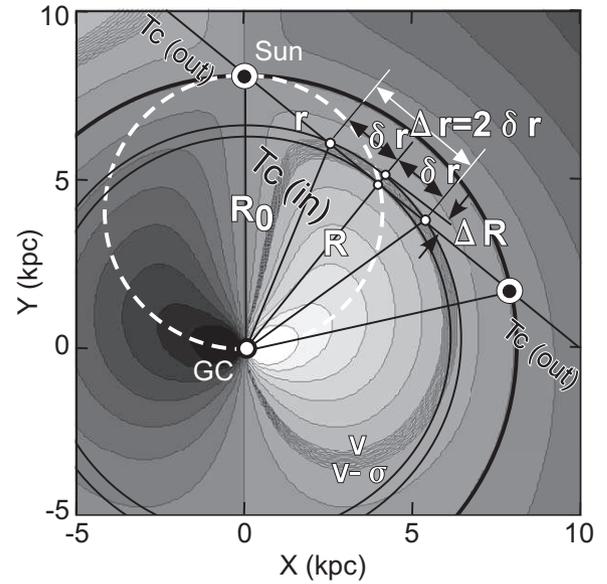} \\
\hskip 15mm \includegraphics[width=6cm]{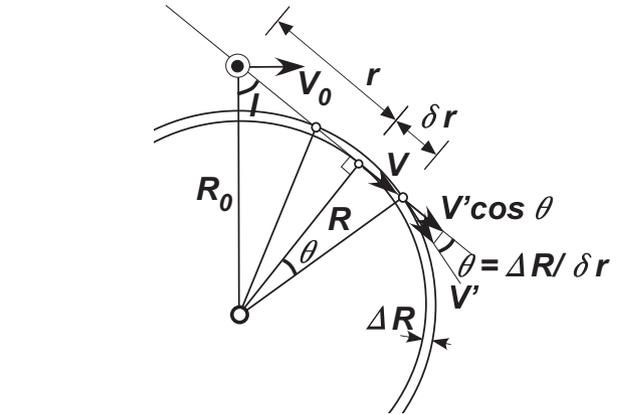}  
\end{center}
\caption{Schematic radial velocity field of the Milky Way. The line of sight passes the velocity-degenerate region (VDR) near the tangent point. Geometrical and kinematic relations among $dR$, $dr$, $\theta$, $\sigv$ and $V$ are shown in the bottom panel. Continuum disk emission is divided into $\Tc({\rm in})$ and $\Tc({\rm out})$ inside and outside the solar circle, and used to calculate the continuum background by $\Tc=\Tc({\rm out})+\Tc({\rm in})/2=[\Tc({\rm obs})+\Tc({\rm obs+180\deg})]/2$. }
\label{vfield}
	\end{figure}

We consider lines of sight through the tangent points (TP) of the galactic rotation in the disk inside the solar circle as shown in figures \ref{vfield} and \ref{armCandW}. The sight line depth $\Delta r$  is related to $r$ displacement from TP, $\delta r$, by $\Delta r=2 \delta r$. Here, $\delta r$ runs through the shadowed region in figure \ref{vfield}, which covers an area having radial velocity between $v$ and $v+\sigv$, where $\sigv$ is the velocity dispersion of the gas. Rewriting as $\sigv=\delta v$, the last factor in equation (\ref{eqtau}) is written as
\begin{equation}
{\Delta v \/ \Delta r} ={1\/ 2} |{\delta v \/ \delta r}|={1\/ 2}|{dv\/ dr}|.
\end{equation} 

\subsection{Kinematic and Geometric formulation} 

We now recall that the radial velocity is related to the galactic rotation by
\begin{equation}
v=V \cos\ \theta-V_0\ \sin \ l
=\left(V {R_0\/ R}-V_0\right)\sin\  l,
\label{v}
\end{equation}
where $\theta$ is the angle between the line of sight and the rotation direction,  $R=(r^2+R_0^2-2r R_0\ {\rm cos} \ l)^{1/2}$ is the galacto-centric distance, $r$ is the distance along the line of sight, and V is the rotation velocity at $R$. 

Hereafter, we discuss the regions near the tangent points, and re-define the galacto-centric distance $R$ by $R=R_0\ |\sin\ l |$.
 Differrentiating $v$ by $r$ as $dv/dr={dv \/ dR}{dR \/ dr}$, we obtain 
\footnote{In our recent paper (equation (5) of Sofue (2017b)) for the 3-kpc HI hole, the factor related to $dR/dr$ was too simplified by geometric effect of rotation velocity, hence the density was over-estimated. The present expression is more correct. }.
\begin{equation}  
{\Delta v \/ \Delta r}\simeq 
{1\/ 2} {V\/ R} {dR \/ dr}\left(1-\dlnv \right).
\label{dvdr}  
  \end{equation}  
Let $\Delta R $ and $\delta r$ be the displacement in $R$ and $r$ corresponding to the radial velocity variation of $\delta v = \sigv$ (figure \ref{vfield}). We consider a near-TP region and assume that $\sigv \ll V$. 
Then, the small lengths $\Delta R$ and $\delta r$ are related to the angle $\theta$ in equation (\ref{v}) as  
\begin{equation}
{dR\/ dr}\simeq {\Delta R\/ \delta r} \simeq \theta.
\label{dRdr}
\end{equation}
Using equation (\ref{v}) we can relate $\theta$ to the variation of radial velocity as  
\begin{equation}
\delta v=\sigv=V(R+\Delta R) \cos\ \theta-V(R).
\end{equation}
Approximating as $V(R+\Delta R)\simeq V(R)+{dV\/dR} \Delta R$ and $1-\cos \theta \simeq \theta^2/2$ for small $\Delta R$ and $\theta$, we have
\begin{equation}
\sigv \simeq {1\/2} \theta^2\ V - {dV\/dR} \Delta R.
\end{equation}
 
Equation \ref{dRdr} is then rewritten as
\begin{equation}
{dR \/ dr} \simeq \sqrt{2\sigv'\over V}, 
\label{eqsigvdash}
\end{equation}
where 
\begin{equation}
\sigv'=\sigv+{dV\over dR} \Delta R
\label{sigvdash}
\end{equation}
is the modified velocity dispersion. 
Here, $\Delta R$ can be approximated as 
\begin{equation}
\Delta R \simeq 2R\sigv/V, 
\label{DeltaR}
\end{equation}
where we used the relation $\theta \simeq \delta r/R$ (figure \ref{vfield}) besides equation (\ref{dRdr}). Thereby, we neglected the second order terms, so that the second term of equation (\ref{sigvdash}) is sufficiently smaller than the first, or $\sigv'\sim \sigv$. 
We now obtain
\begin{equation}
\sigv'\simeq \sigv\left(1+2 \dlnv \right).
\label{sdashdlnVdlnR}
\end{equation}
This relation is valid as the the second term is small enough, because the derivative is logarithmic and is obtained as an averaged slope in $\Delta R \sim 0.7 (R/R_0)$ kpc. In fact, in the middle disk at $R\sim 4$ kpc, $\Delta R\sim 0.35$ kpc and the averaged slope is $dV/dR\sim 3$ \kms kpc$^{-1}$, yielding $dV/dR \Delta R\sim 1$ \kms, much smaller than $\sigv=10$ \kms.

Equation (\ref{eqtau}) can be now written as
\begin{equation} 
\tau(v)\simeq{n  \/ \Xhi {\Ts}}\left[{1\/\sqrt{2}}{ \sqrt{\sigv' V}\/R} \left(1 - \dlnv \right) \right]^{-1}.
\label{eqtaudvdr}
\end{equation}
We apply this expression to solve equation (\ref{eqTb}) for the unknown parameters $\Ts$ and $n$ in a fixed $R$ (longitude) range of width $\Rwidth=\Rmax-\Rmin$, where $\Rmax$ and $\Rmin$ are the outer and inner radii of the ringlet (small circular arc within $\theta$ from the TP in figure \ref{vfield}) along the tangent-point circle. For later convenience, we rewrite the equation for $n$ as
\begin{equation} 
n\simeq {1\/\sqrt{2}}\tau(v) \Xhi \Ts {\sqrt{\sigv' V}\/ R} \left(1-{\dlnv} \right)  .
\label{eqdendvdr}
\end{equation} 

\begin{figure} 
\begin{center}     
\includegraphics[width=8cm]{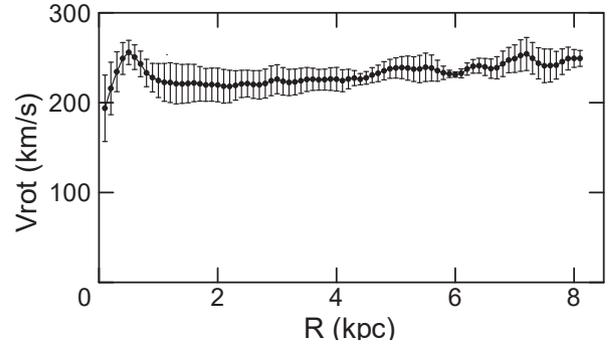}   
\end{center}
\caption{HI rotation curve obtained by using LV diagrams from the LAB and GASS HI line surveys. 
}
\label{rc}
	\end{figure}

Figure \ref{rc} shows the rotation curve of the Galaxy as obtained by tracing the terminal velocities in the longitude-velocity (LV) diagrams from LAB and GASS HI survey data. This curve is used in the following analyses to calculate $\dlnv$ in the above equations after smoothing by a Gaussian function with a half width of $\Delta R$.  

We also note that equations (\ref{eqtaudvdr}) and (\ref{eqdendvdr}) fail near the Sun at $R>R_0-\Rwidth'$, where $\Rwidth'\simeq R_0 \sigv'/V_0$, because the observer is inside the radial ring within which the fitting is performed (shaded area in figure \ref{vfield}). 
So, we do not use data at such radii, and instead, employ the local values from our previous paper (Sofue 2017a). 
        
\subsection{Optically-thin case }

If the gas is optically thin, we have $\tau \simeq \Tb/(\Ts-\Tc)$, and then
\begin{equation}
n\simeq {1\/\sqrt{2}}\Xhi \Tb {\Ts\over \Ts-\Tc} {\sqrt{\sigv' V}\/ R} \left(1-\dlnv \right).
\label{eqthin}
\end{equation}
If the rotation curve is flat, we obtain an approximation as
\begin{equation}
n\sim {1\/\sqrt{2}}\Xhi \Tb {\Ts\over \Ts-\Tc}{\sqrt{\sigv V}\/ R}.
\label{eqthinflat}
\end{equation}
If we can neglect the background continuum, we obtain a very rough estimation as
\begin{equation}
n\sim {1\/\sqrt{2}}\Xhi \Tb{\sqrt{\sigv V}\/ R}.
\label{eqthinflat}
\end{equation}
This yields an approximate density of $n\sim 0.25\ (R/8 {\rm kpc})^{-1}\ {\rm H\ cm}^{-3}$ for $V\sim 238$ \kms, $\Tb\sim 100$ K and $\sigv \sim 10$ \kms.
However, it must be remembered that, unless the gas has sufficiently high spin temperature compared to $\sim 100$ K, this approximation causes significant under-estimation. In fact, we later show that the under-estimation amounts to a factor of $\sim 4$ for the cold HI gas.

\section{Data}

\subsection{HI cubes}

We used the LAB and GASS HI surveys to make longitude-velocity (LV) diagrams along the galactic plane ($b=0\deg$). On the diagrams we traced short LV ridges nearest to the terminal velocities at various longitude ranges using a data-reading software on MAKALII (a software for image processing for SUBARU). This yielded many sets of ($l,v$) values in each of the longitude ranges, which had span of of $\sim 2\deg$ to $10\deg$.

\begin{figure} 
\begin{center}      
\includegraphics[width=8cm]{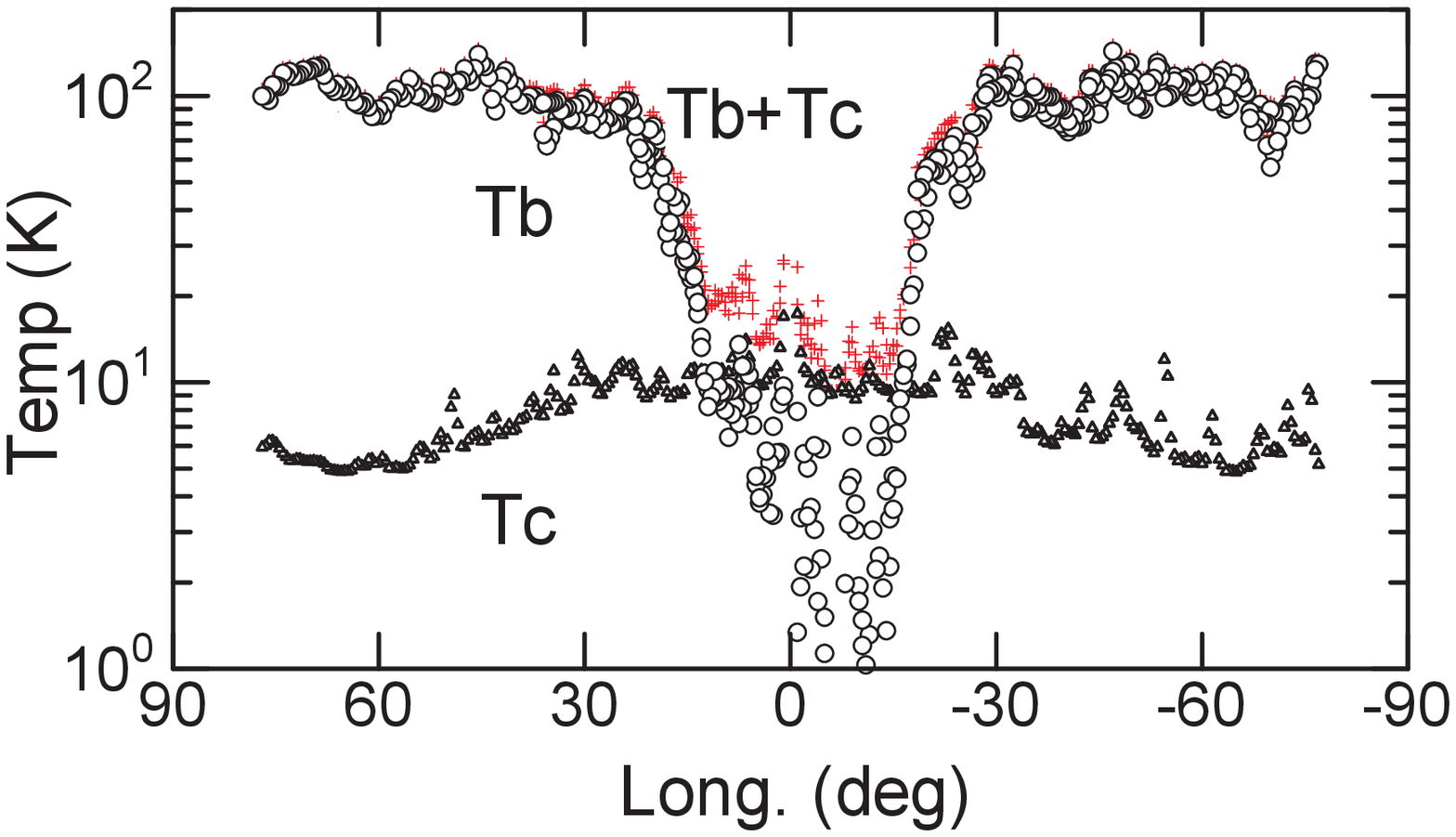}   
\includegraphics[width=8cm]{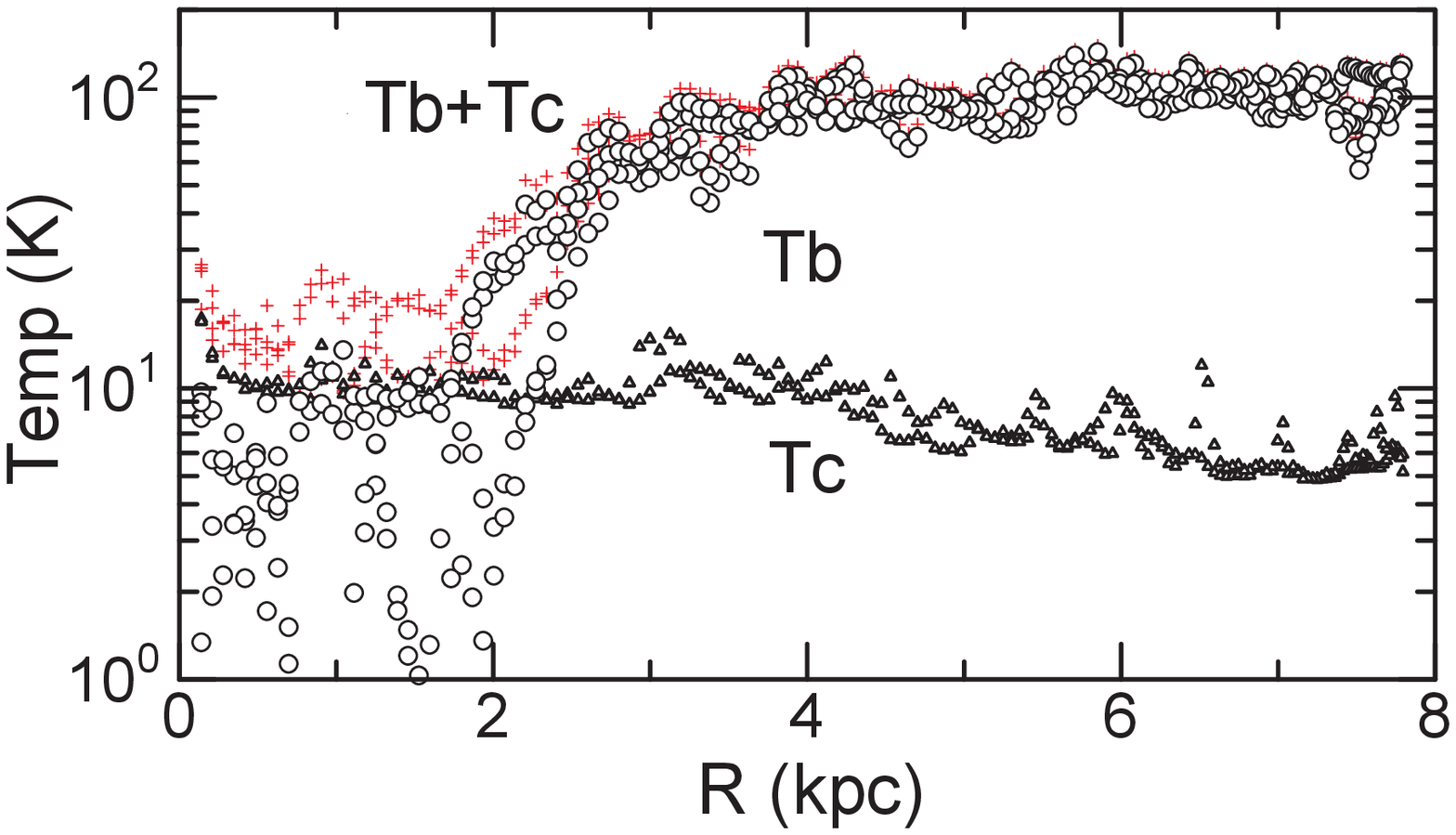}   
\end{center}
\caption{HI brightness temperature $\Tb$ at tangent points (open circles), background continuum brightness temperature at 1420 MHz $\Tc$ given by equation (\ref{TcBack}) (triangles), and their sum $\Tb+\Tc$ (crosses) giving the lower limit to $\Ts$ for optically thick case. Top panel shows a plot against longitude, and bottom against galacto-centric distance of the tangent point. 
}
\label{TbTc}
\begin{center}       
\includegraphics[width=8cm]{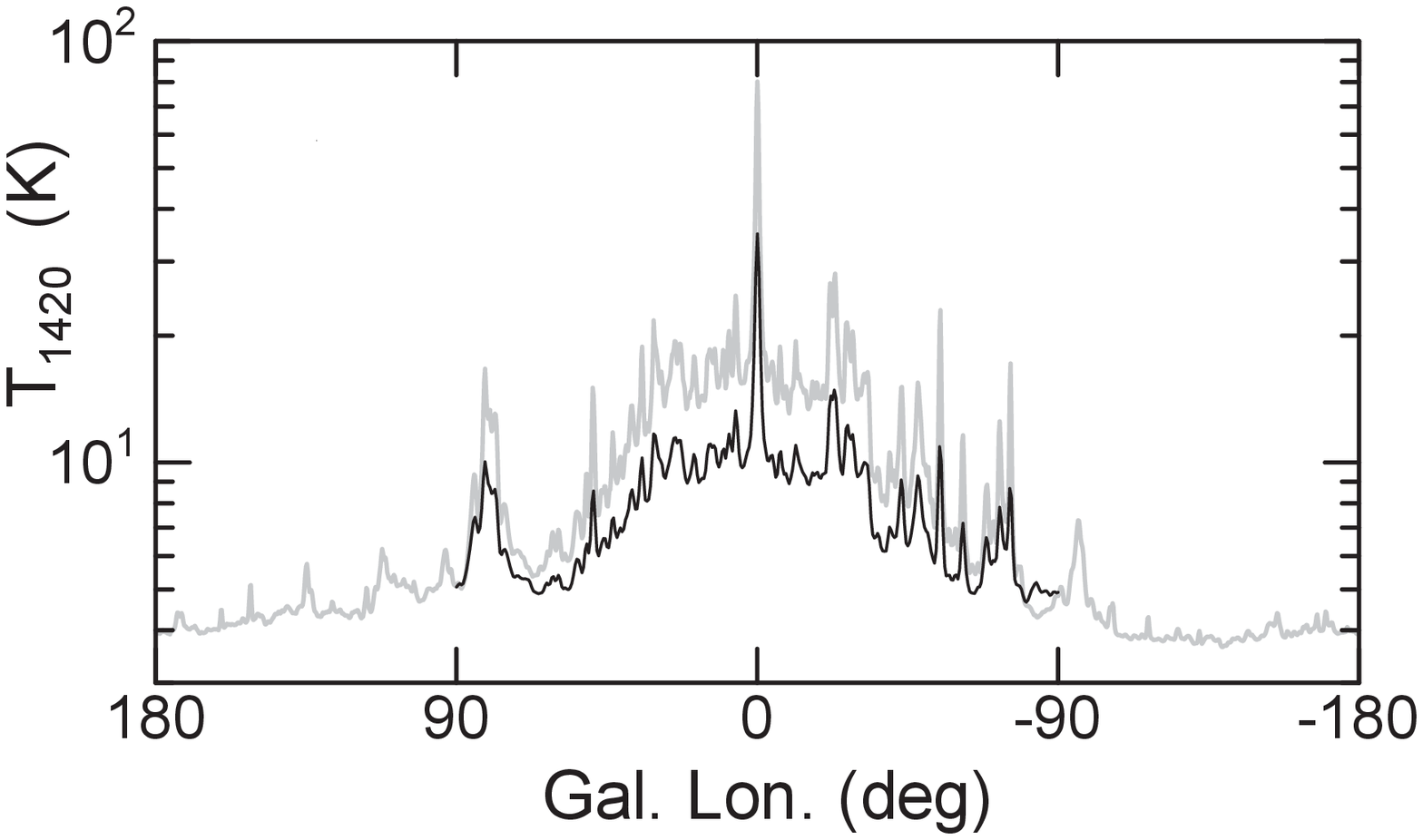}   
\end{center}
\caption{[Grey line] 1420 MHz continuum absolute brightness temperature along the galactic plane from the Stockert-Villa-Elisa all sky survey (Reich et al. 2009). The cosmic micro-wave background (CMB) is included. 
[Black line] Background emission beyond tangent points calculated by equation \ref{TcBack}. 
}
\label{Tc}
	\end{figure}

Thus obtained LV ridges were overlapped with the neighboring ridges near their edges. In order to avoid redundancy for this overlapping, we took Gaussian-running average with a full width of half maximum of $0\deg.6$ (beam width for LAB and continuum) around each longitude mesh at interval of $0\deg.5$. Finally we obtained tables of regridded $(l,v)$ at longitude interval of $0\deg.5$ at $-90\deg \le l \le +90\deg$ from LAB, and at $-90\deg \le v \le +35\deg$ from GASS data. Table \ref{tabchisq} lists the adopted input parameters. Figure \ref{TbTc} (top) shows the thus obtained $\Tb$ along the galactic plane. 
         
	\begin{table} 
\caption{Input parameters for least-$\chi^2$ search} 

\begin{tabular}{ll}  

\hline 
\hline   
Long. grid of LVD from LAB  & $0\deg.5$ \\ 
ibid  GASS   & $0\deg.2$ \\ 
ibid  SVE 1420 MHz & $0\deg.25$ \\ 
Regridded interval for analysis & $0\deg.5$ \\
Half width for running average &  $\pm 0\deg.3$\\

$R$ interval$^\dagger$ $\Rwidth$ &  0.1 kpc\\
Search range about $R$ &  $\pm 0.1$ kpc\\

Search meshes in $(n,\Ts)$ plane &$100\times 100$\\
Search range of $n$ (Min/Max) &  0.03 / 30 \Hcc\\
search range of $\Ts$ (Min/Max) &   10 / 10000 K\\
Search dex interval &  1.07152 =(max/min)$^{1/100}$\\
\hline
\end{tabular}  
$^\dagger R=R_0\ | \sin \ l|$ is the galacto-centric distance along the tangent-point circle. 
\label{tabchisq} 
	\end{table}

\subsection{1420 MHz continuum map}

Radio continuum brightness temperature at 1420 MHz was read from the SVE all-sky survey (Reich et al. 2001). The gray line in figure \ref{Tc} shows the brightness distribution, $\Tc$(obs), along the galactic plane, which includes the cosmic background radiation (CMB).

We assume that the continuum emission coming from the space beyond the solar circle, $\Tc({\rm out})$, is equal to that in the opposite direction of the Sun, $\Tc({\rm obs}+180\deg)$, at longitude $l+180\deg$ (figure \ref{vfield}). We also assume that the galactic contributions between the Sun and TP ($\Tc({\rm in})/2$)  and between TP and far-side point on the solar circle ($\Tc({\rm in})/2$) are equal. We thus estimate the background emission beyond TP by
\begin{equation}
\Tc={1 \/ 2} \Tc({\rm in})+\Tc({\rm out})={1\/2}[\Tc(\rm obs)+\Tc({\rm obs}+180\deg)],
\label{TcBack}
\end{equation}
where we used $\Tc({\rm in})=\Tc({\rm obs})-\Tc({\rm obs}+180\deg)$. 
The lower black line in figure \ref{Tc} shows the thus calculated $\Tc$. Note that the contribution of $\Tc$ to $\Ts$ amounts to $\sim 5-12$ K in the inner galactic disk.

In order to apply $\chi^2$ analysis for the emissions at TP at various galacto-centric distances $R$ in the next section, we further convert the longitudes to by $R=R_0 |\sin\ l|$. The lower panel of figure \ref{TbTc} shows the distributions of $\Tb$ and $\Tc$ against $R=R_0\ |\sin \ l|$. We also indicate $\Tb+\Tc$ as measures of lower limits of expected $\Ts$ at individual points.

\section{Determination of $\Ts$ and $n$ by the Least-$\chi^2$ Method}
\label{sectionresult}

\subsection{The method}
 
We solve equation (\ref{eqTb}) for the two  unknown parameters, ${\Ts}$ and $n$. Input observables are $\Tb$, $\Tc$, $V(R)$ and $\sigv$ at individual positions in each search range from $R=\Rmin$ to $\Rmax$. We measure $\Tb$ using the LAB and GASS HI surveys, $\Tc$ using SVE 1.4 GHz survey, $V(R)$ from the terminal velocities using LV diagram, and $\sigv$ from the literature.  

\begin{figure} 
\begin{center}       
\includegraphics[width=6cm]{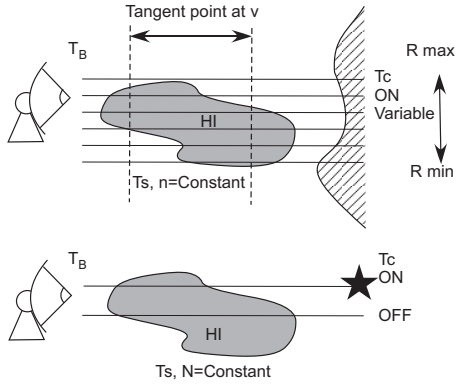}         
\end{center}
\caption{Illustration of the present method (top), compared with the current absorption/emission (on-off) method (bottom). }
\label{methodOnOff}
	\end{figure}

Mathematically, and only if the errors are sufficiently small, the equation can be solved by knowing two sets of these observables at two different positions. Practically, however, we use the equations to estimate the most probable $\Ts$ and $n$ by the least-$\chi^2$ method for many sets of the observables in the considered region. 
The present method may be compared to the classical on (absorption) - off (emission) method, as illustrated in figure \ref{methodOnOff}.   

 We assume that the gas along each line of sight is dominated either by cold or warm HI gas, so that the search for the least $\chi^2$ shall automatically select one of the two possible solutions, either cold or warm. In principle, a case where the two components are mixed on the same line of sight could be solved by iteration, starting with one component, either cold or warm, then the residual by another component. Obviously this needs a much high-quality and larger scale data as well as more careful transfer of the HI line, and would be a subject for the future.

The dispersion velocity of HI gas has been measured to be $\sigv\simeq 7-10$ \kms (e.g., Liszt 1983). Similar dispersion of $\sim 12$ \kms have been obtained in many spiral galaxies (Mogotsi et al. 2016). Recently, Marasco et al. (2017) determined $\sigv$ precisely using spectral fitting to terminal velocity profiles, and obtained $\sigv\sim 8$ \kms near the Sun, stays almost constant at $\sim 10$ \kms in the mid-disk, gradually rises to $\sim 12$ \kms at $R\sim 2$ kpc, and further toward the GC. Following this work, we here assume a constant dispersion of $\sigv=10$ \kms. Dependence of the result on $\sigv$ will be discussed later in detail.

We define $\chi^2$ by
\begin{equation}
\chi^2=\Sigma_i \left[\Tb(i) -\Tb(i;{\rm cal}) \/ \sigma_{\Tb}\right]^2,
\label{eqchisq}
\end{equation}
where $\Tb(i)$ is the observed $\Tb$ at various ($i$-th) longitudes, and $\Tb(i;{\rm cal})$ is the calculated $\Tb$ using equation (\ref{eqTb}) for observed $\Tc$ at the same longitudes. Since the LV data do not present errors at individual positions, we approximate them by the dispersion $\sigma_{\Tb}$ of the observed $\Tb$ values in each $R$ range. We neglect the effect caused by the dispersion of $\Tc$, because $\Tc$ and its dispersion are an order of magnitude smaller.

The $\chi^2$ values were calculated at all points on the $n-\Ts$ space divided into $100\times 100$ meshes in logarithmic dex interval ($1000^{1/100}$ times) from $n=0.03$ to 30 \Hcc and from $\Ts=10$ to 10000 K. 

The least-$\chi^2$ value and the best-fit $n$ and $\Ts$ were calculated by fitting a parabola to three points in each direction of $n$ and $\Ts$ around the mesh point giving the smallest $\chi^2$.
The errors of the best-fit values were calculated using this parabola as displacements of $n$ and $\Ts$ from the best-fit values that yielded 
increase of $\chi^2$ (not $C^*$) by 1 from the minimum $\chi^2$. This means that the calculated $\Tb$ lies within $\delta \Tb \sim \sqrt{1/N}\sigma_{\Tb}$ from the true value, where $N$ is the number of data points in the $R$ bin.

\subsection{Global behavior of $\chi^2$ distribution}

\begin{figure} 
\begin{center}  
\includegraphics[width=6cm]{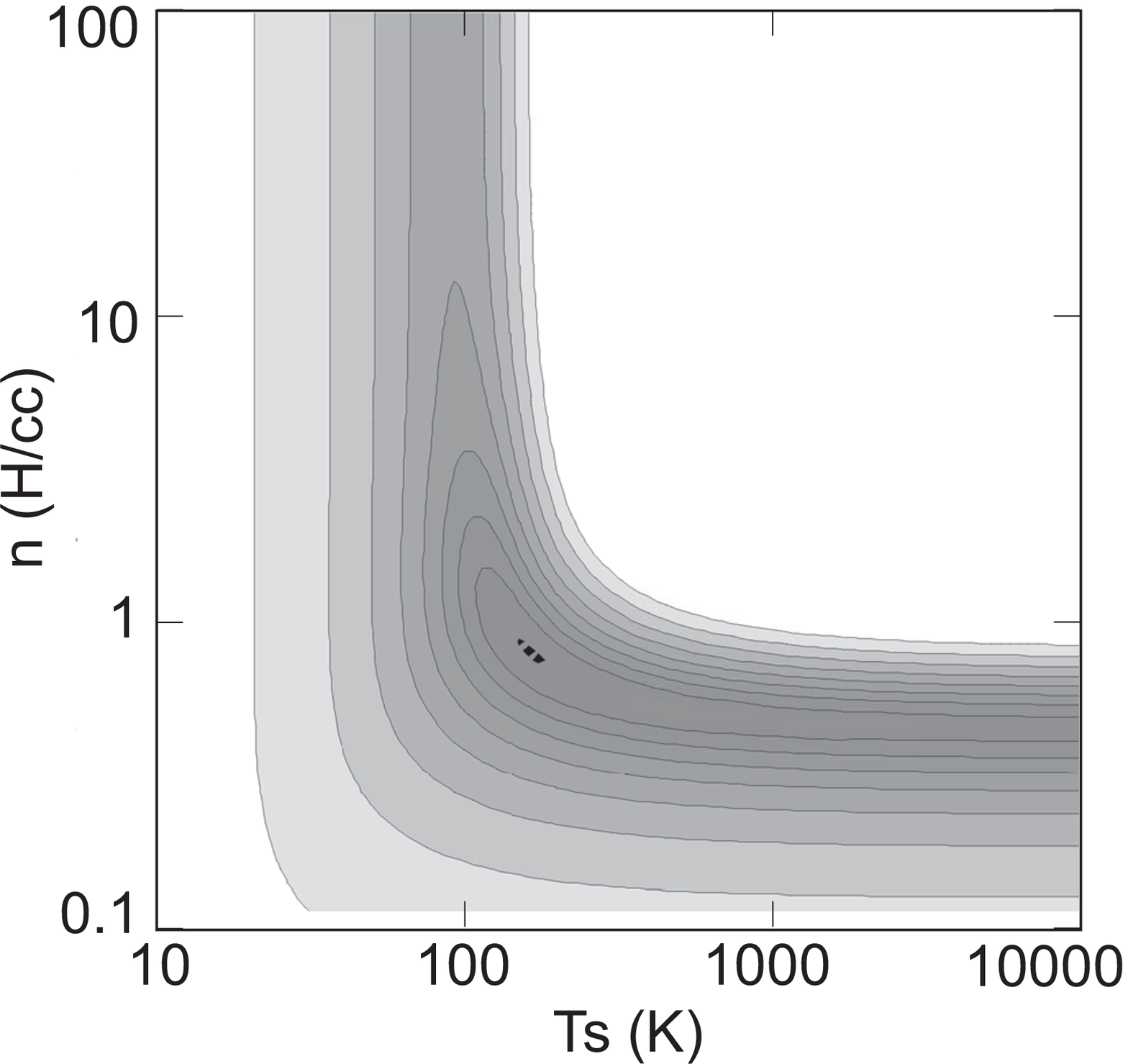}  
\end{center}
\caption{Distribution of $C^*=\chi^2/{\rm Min}(\chi^2)$ in the $\Ts-n$ plane using all the disk data from $R=0$ to 8 kpc. Contours are drawn tightly from 1.0 to 1.05, and then every $\sqrt{2}$ dex interval. The most probable values are in the darkest region.}
\label{cmap0to8} 
          
\begin{center}  
\includegraphics[width=6cm]{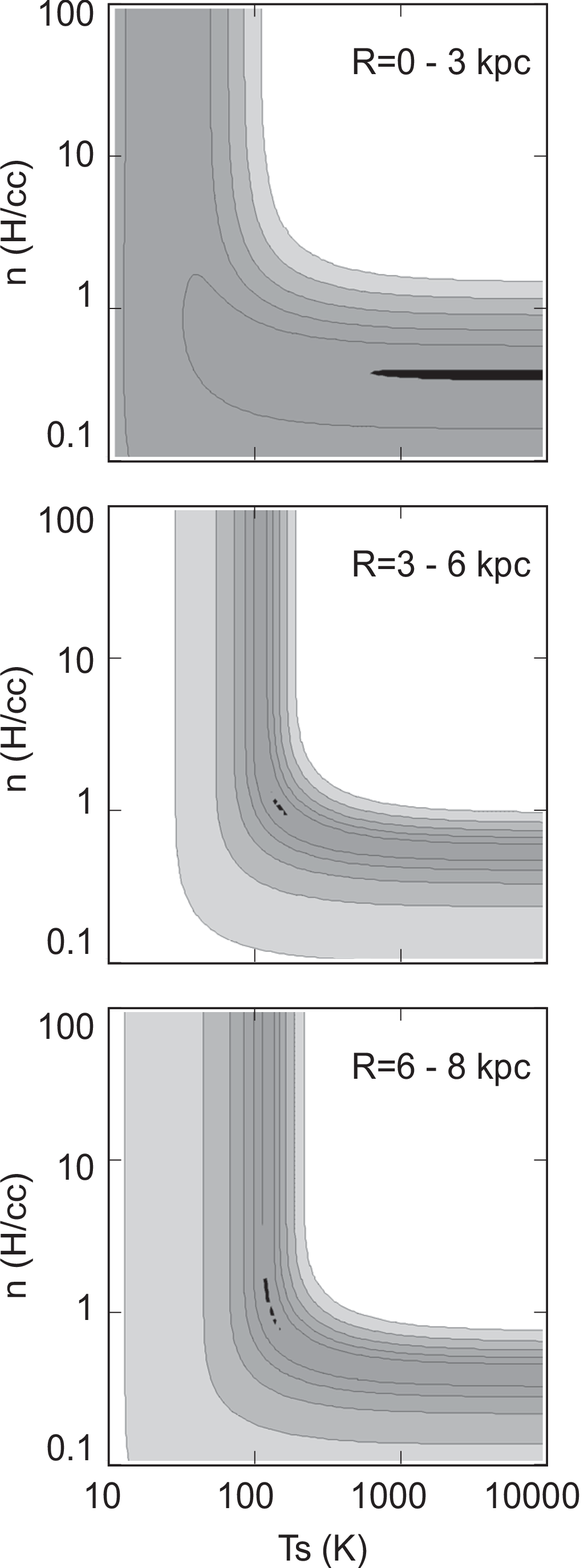}  
\end{center}
\caption{Same as figure \ref{cmap0to8}, but for three regions at $R=0$ to 3 kpc (typically for WNM), 3 to 6, and 6 to 8 kpc (almost CNM). Contour interval is here 2 dex.  }
\label{c3maps} 
	\end{figure}

Figure \ref{cmap0to8} shows the distribution of calculated $C^*=\chi^2/{\rm Min}[\chi^2]$ in the ($\Ts,n$) plane by contour diagrams obtained for all the data in figure \ref{TbTc} for $\Rwidth=8$ kpc. Namely, it shows averaged $C^*$ in the Galactic disk inside the solar circle. The $\chi^2$ minimum appears at $(n,\Ts)\sim (1.3,120)$ (H cm$^{-3}$, K) as the averaged density and spin temperature in the inner disk of the Galaxy. The spin temperature is consistent with the widely adopted value of $\Ts \sim 130$ K (see the literature in section 1), while the density is about three times the currently obtained (e.g., Burton 1976; Marasco et al. 2017). 

We then divide the disk into three wide rings from $R=0$ to 3 kpc ($\Rwidth=3$ kpc), 3 to 6, and from 6 to 8 kpc ($\Rwidth=2$ kpc), within each of which $n$ and $\Ts$ are assumed to be constant. Figure \ref{c3maps} shows the $\chi^2$ distributions. In the innermost disk, the best-fit spin temperature appears at $\Ts\ge \sim 2000$, although $\chi^2$ minimum is rather broad. On the other hand, the density is determined better at around $n \sim 0.3$ \Hcc. Thus, the HI gas inside 3 kpc ring is warm (WNM) and optically thin.

In the main disk at $R>\sim 3$ kpc, the best-fit $\Ts$ appears at $\sim 100-150$ K, indicating that the gases are in the CNM state. The density, $n\sim 1-2$ \Hcc, is again higher than the currently known values based on the optically thin assumption.

\subsection{Variations of $\Ts$ and $n$ as a function of $R$}
\label{secresult}

\begin{figure}
\begin{center}     
\includegraphics[width=8.5cm]{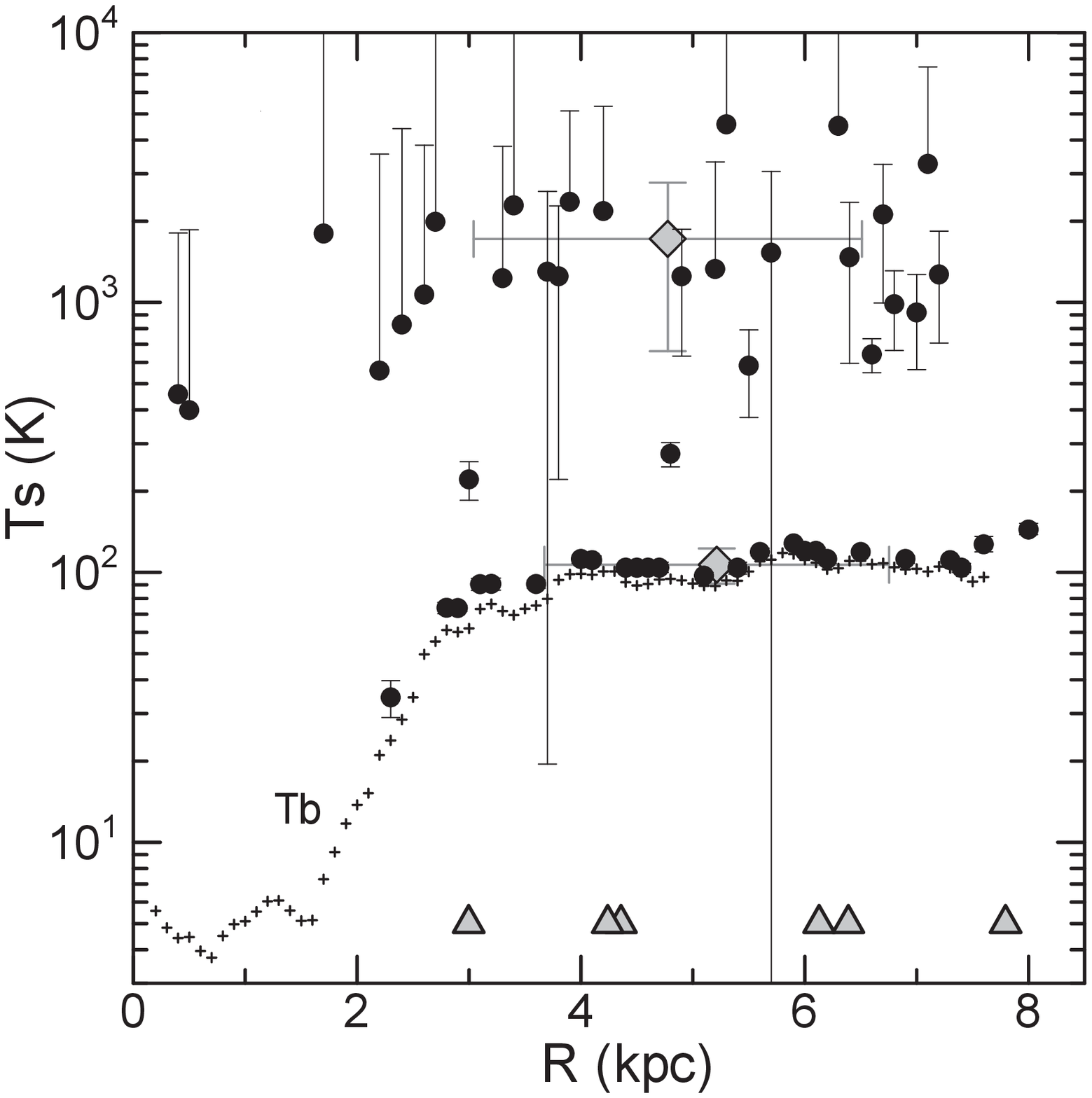}   
\end{center}
\caption{
The best-fit $\Ts$ (filled circles) and observed $\Tb$ (crosses) for $\sigv=10$ \kms plotted against $R=R_0 \ |\sin \ l|$ (galacto-centric distance). Error bars indicate ranges allowing increase of $\chi^2$ by 1 from the minimum. Up-sided bars indicate errors greater than the fitted values, showing less reliable values.  Big diamonds are averages (regardless errors) and standard deviations of the plotted values between $R=3$ and 8 kpc separately for cold ($\Ts \le 200$ K) and warm HI ($> 300$ K). Big triangles indicate tangential directions of the galactic rings and spiral arms. 
}
\label{fit_Ts}  

\begin{center}       
\includegraphics[width=8.5cm]{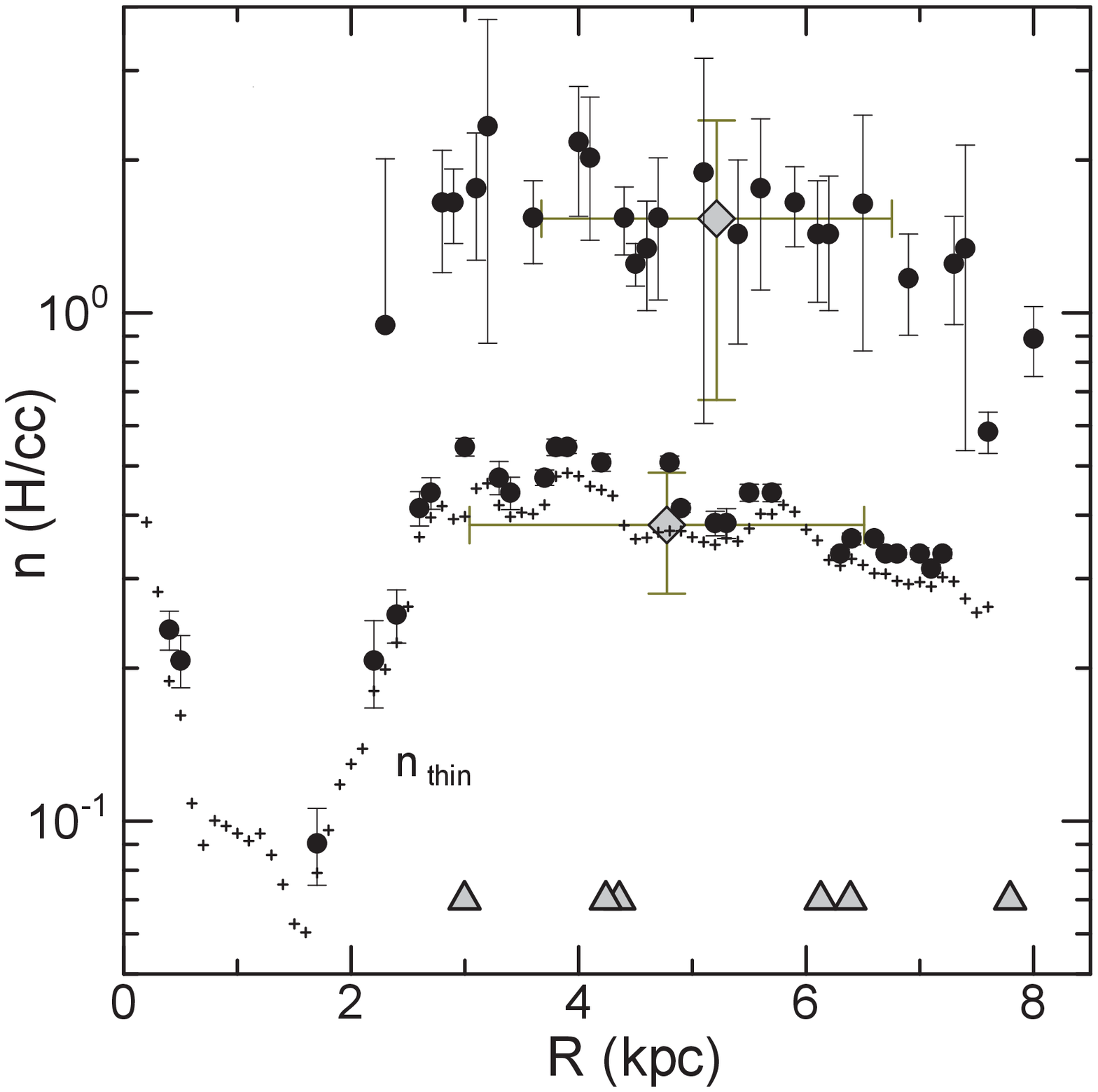}  
\end{center}
\caption{  
The best-fit $n$ (circles) and $n_{\rm thin}$ for optically thin limit (crosses).  Error bars, big diamonds, and triangles are as for figure \ref{fit_Ts}.  
} 
\label{fit_n}  
\end{figure}

\begin{figure}
\begin{center}     
\includegraphics[width=7cm]{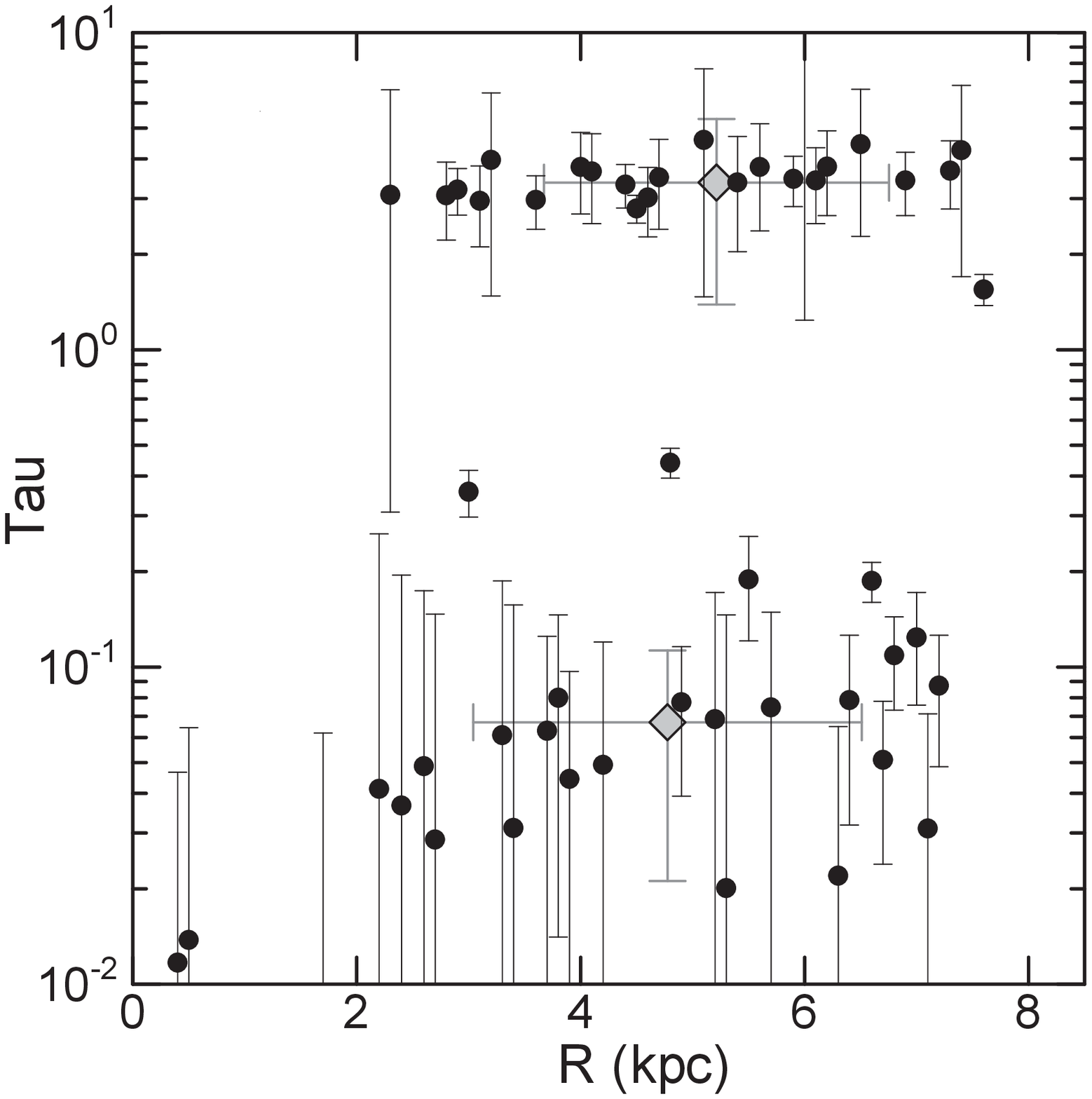}  
\includegraphics[width=7cm]{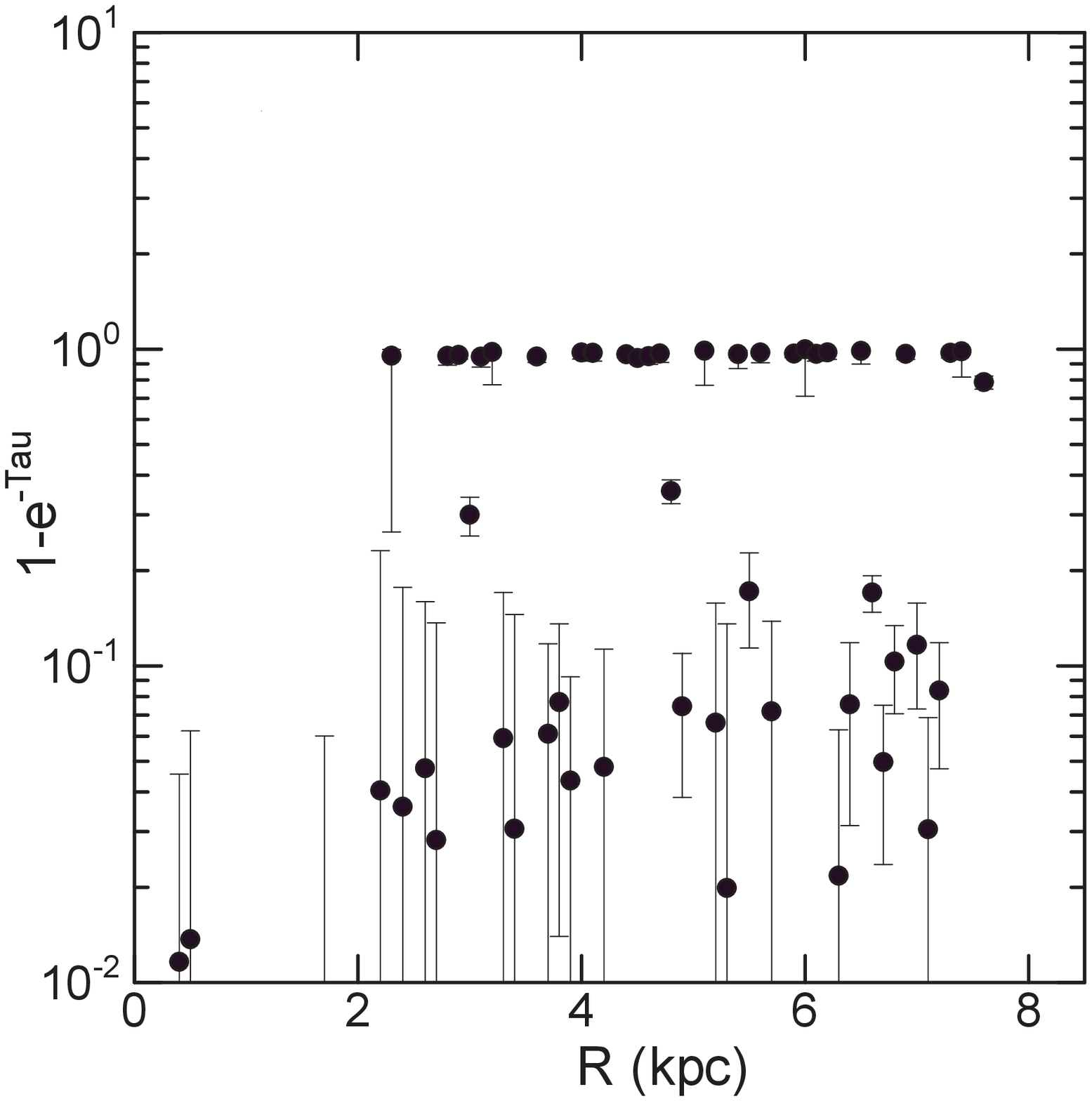}    
\end{center}
\caption{ 
Same as figure \ref{fit_Ts}, but for the optical depth $\tau$ (top) and $1-e^{-\tau}$ (bottom).   
}
\label{fittau}
\end{figure}

In order to investigate the variation of $\Ts$ and $n$ as a function of $R$ in more detail, we divide the galactic disk near the tangent-point circle into a larger number of ringlets  of width $\Rwidth= 0.1$ kpc, and apply the least-$\chi^2$ search in each bin. Figures \ref{fit_Ts}, \ref{fit_n} and \ref{fittau} show plots of the best-fit values of $\Ts$, $n$, $\tau$, and $1-e^{-\tau}$ against  $R$. During the search, a few points showing $\Ts$ greater than 8000 K were rejected as unrealistic values, because the gas should not be neutral beyond such high temperature. Error bars indicate ranges of the fitted values that allow for increase of $\chi^2$ by 1 from the minimum. Up-sided bars indicate errors greater than the fitted values, and hence less reliable results. 

 Big gray diamonds are averages (regardless the errors) and standard deviations of the plotted values between $R=3$ and 8 kpc, separately calculated for cold ($\Ts \le 200$) and warm ($ > 300$) HI. Considering the two major crowding of $\Ts$ plot in figure \ref{fit_Ts}, we rejected two isolated points at $\sim 250$ K from averaging.

\subsubsection{Spin temperature}

Figure \ref{fit_Ts} shows that the HI gas in the main disk at $R\ge \sim 3$ kpc has either lower temperature at $\Ts\sim 100-120$ K, representing cold HI, or higher at $\Ts\sim 1000-3000$ K, representing warm HI.
The warm and cold HI exactly correspond to low and high densities, respectively, and hence, low and high optical depths. The averaged $\Ts$ over $3\ge R \ge 8$ kpc is $\Ts=106.7\pm 16.0$ K for cold HI, and $\Ts=1721\pm 1060$ K for warm HI (table \ref{tabav}). 
The figure shows that $\Ts$ of warm HI is largely scattered around the mean, whereas $\Ts$ of the cold HI stays almost constant.

These temperatures may be compared with the current measurements using the ON (absorption)-OFF (emission) spectral method for the local HI gas and clouds toward radio continuum sources. 

The here derived $\Ts\simeq 106 K$ for cold HI is in the range of $40-200$ K as derived for the CNM (Roy et al. 2003b; also the literature in section 1), and is consistent with the column-density weighted mean value of $\Ts = 99-108$ K obtained for the local HI gas (Heiles and Troland 2003a,b). Kanekar et al. (2011) plotted $\Ts$ measured toward quasars as a function of galactic latitude, showing a clear decrease of $\Ts$ toward the galactic plane, where two near-plane sources at $b=-1\deg.60$ and $+2\deg.56$ indicated $\Ts=89$ and 119 K, respectively, which agree with the present estimate at the galactic plane ($b=0\deg$). 
The results may be also compared with the large number current measurements, suh as by Begum et al. (2010) showing $\Ts\sim 100$ K for a dozen of cold HI clouds; 
Gibson et al. (2000) $\Ts=40-70$ K; 
Crovisier et al (1981) $ 60-200$ K;
Albinson et al. (1986) $\sim 80$ K;
Mebold et al. (1982) $\sim 86$ K (median value);
and Liszt et al. (1983) who showed $Ts=40(1-e^{-\tau})^{-0.5}$ K.

The spin temperature for warm HI, $\Ts\sim 1700$ K, is lower than the kinetic temperature of $\sim 5000-8000$ K required for thermally stable interstellar clouds (Roy et al. 2003b). It lies rather in the intermediate temperature range expected to exist as a transient phase (Field 1969). We will later show that the presently detected warm HI gas is distributed in the thermally unstable region on the phase diagram (subsection \ref{ISM}).
The result for warm HI may be further compared with those by 
Mebold et al. (1982) giving $200-1000$ K in cloud envelopes; 
Dwarakanath,  et al. (2002) indicating $\Ts=3600\pm 360$ in local HI gas;
and continuously increasing $\Ts$ from 40 to $\sim 2000$ K for low-$\tau$ clouds obeying the above relation by Liszt et al. (1983).

\subsubsection{Density} 
The HI density $n$ has a central peak in the GC, and decreases steeply toward the HI hole at $R\sim 2$ kpc (Lockman et al. 2016; Sofue 2017b). The density, then, increases toward $R\sim 3$ to 4 kpc, attaining high values around $n\sim 2$ \Hcc. It then gradually decreases outward till $R\sim 8$ kpc, where the values coincide with the local value of 0.9 \Hcc (Sofue 2017a).

The fitted densities at $R>\sim 3$ kpc are divided into two groups, one high density at $n\sim 1-1.5$ \Hcc, and the other low density at $\sim 0.2-0.5$ \Hcc. These two correspond to warm and cold HI, respectively. In the figures we also plot densities calculated on the optically-thin assumption. The density of warm HI approximately follows that calculated for optically thin assumption. 

In more detail, the density of cold HI has peaks at the the 3-kpc expanding ring, the 4-kpc molecular ring, and spiral arms No. 3 and 4 (Nakanishi and Sofue 2006). In contrast to this, warm HI tends to appear in the inter-arm and inter-ring regions.

By averaging the values over the galactic disk between $R=3$ and 8 kpc, we obtained mean (representative) values of $n= 1.53\pm 0.85$ \Hcc for cold HI gas, and $0.38\pm 0.10$ \Hcc for warm (table \ref{tabav}). It should be stressed that the HI density in the CNM phase, and hence the mass of galactic HI gas, has been underestimated by a factor of $\sim 4$ in the current optical thin assumption compared to the value by the present determination.

	\begin{table*} 
\caption{Spin temperature and density averaged at $3 \le R \le 8$ kpc for $\sigv=10$ \kms, and HI masses inside the solar circle.\\}
        \begin{center}
\begin{tabular}{llll}  

\hline 
\hline  
 & Cold HI (CNM)& Warm HI (WNM) &Total \\ 
\hline    
Location\dotfill& Arms and rings & Inter-arm &---\\ 
Density, $n$ (\Hcc)\dotfill & $1.53\pm 0.85$ &  $0.38 \pm 0.10$ &--- \\
Spin temp., $\Ts$ (K)\dotfill & $106.7 \pm 16.0 $  & $ 1721 \pm 1060$ &--- \\  
Optical depth, $\tau$ & $3.37 \pm 1.98$  & $ 0.067\pm 0.046$ &--- \\ 
\hline
HI mass ($R\le 8$ kpc) ($\Msun$)  \dotfill & $\sim  1.47 \times 10^9$ & $\sim 0.32 \times 10^9 $ & $ \sim  1.78 \times 10^9 $ \\
\hline
\end{tabular}  
\label{tabav} 
\end{center}
	\end{table*}   

\subsubsection{Optical depth}

Optical depth $\tau$ is small in the GC region due to the geometrical effect of the small line-of-sight length as well as the large values of $dv/dr$. Hence, the HI gas in the inner-most region at $R<\sim 2$ kpc may be well determined by the optically thin assumption. 

The optical depth increases with the $R$ and tends to have two states, one having large thickness with $\tau \sim 3-4$, and the other with small thickness at $\tau \sim <0.1$. The optically thick and thin regions exactly correspond to the cold and warm HI regions, respectively.
In the figure we also show $1-e^{-\tau}$, which is nearly equal to unity in cold HI regions.

In so far as the velocity dispersion is sufficiently small compared to rotation velocity, the tangential directions of the galactic rotation are velocity-degenerate regions (VDR) as defined by Sofue (2017a). This is the reason why the cold HI showed large optical depth.
        
\subsubsection{Appropriate $R$ intervals}

We comment on the appropriate resolution in the analysis. The present $R$ interval, $\Rwidth =0.1$ kpc and width ($\pm \Rwidth$ kpc) resulted in reasonable fitting in almost all $R$ bins at $R\ge\sim 2$ kpc. However, if we take a smaller interval, e.g. 0.05 kpc, the input data points in each bin become too few for good fitting.
Also for wider intervals, e.g. $\Rwidth \sim 1 $ kpc, the result was also not definite, showing too large scatter due to mixture of warm and cold HI gases having diverging solutions.
However, if we take a much wider interval, e.g. $\Rwidth \sim 3$ to 8 kpc, the input data number increases to reach the global values of $\Ts$ and $n$ rather safely as seen in figures \ref{cmap0to8} and \ref{c3maps}.
        
\subsection{ISM Physics: Correlation among $n$, $\Ts$ and $\tau$}
\label{ISM}

\begin{figure} 
\begin{center}      
\includegraphics[width=7cm]{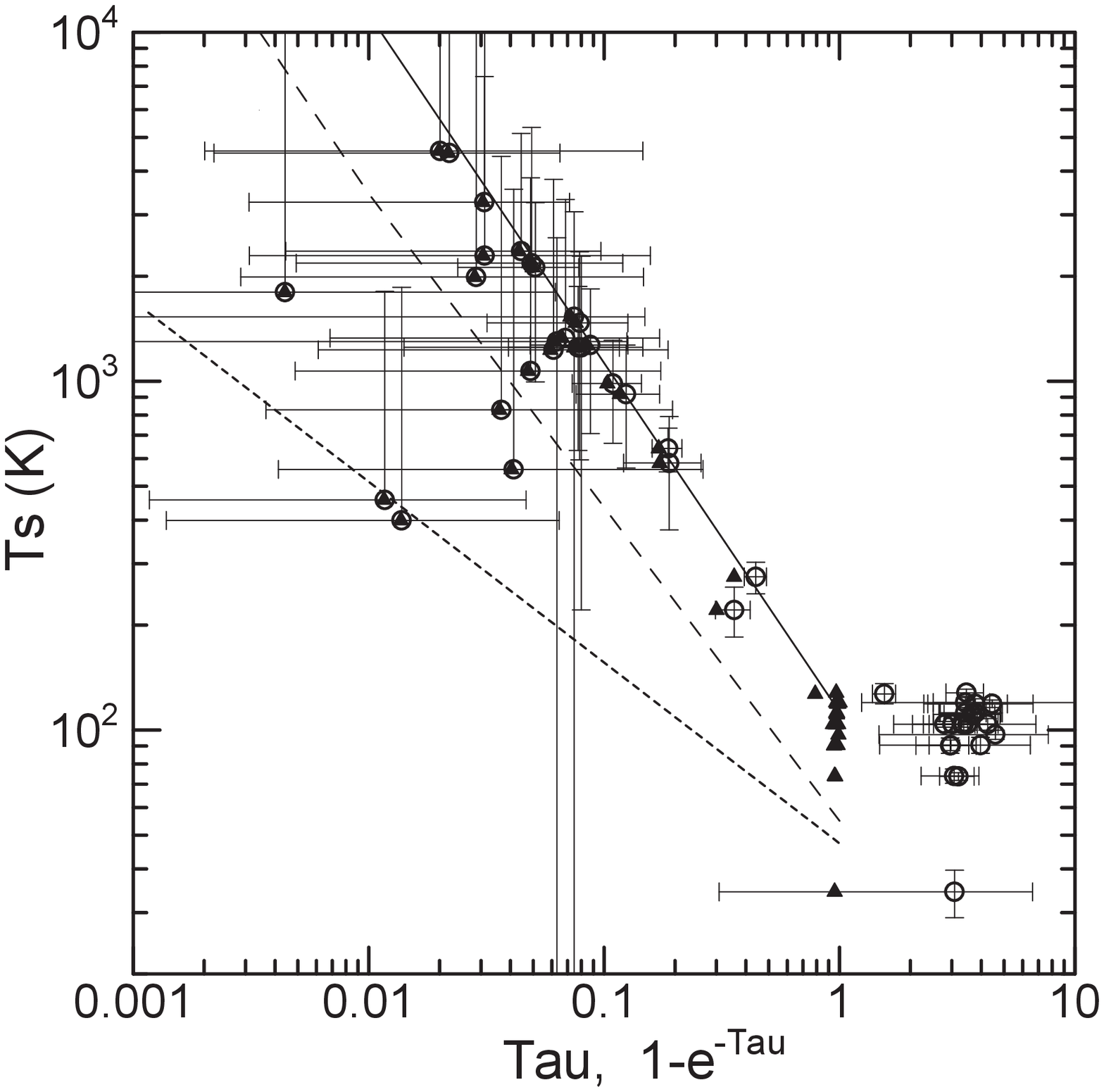}    
\includegraphics[width=7cm]{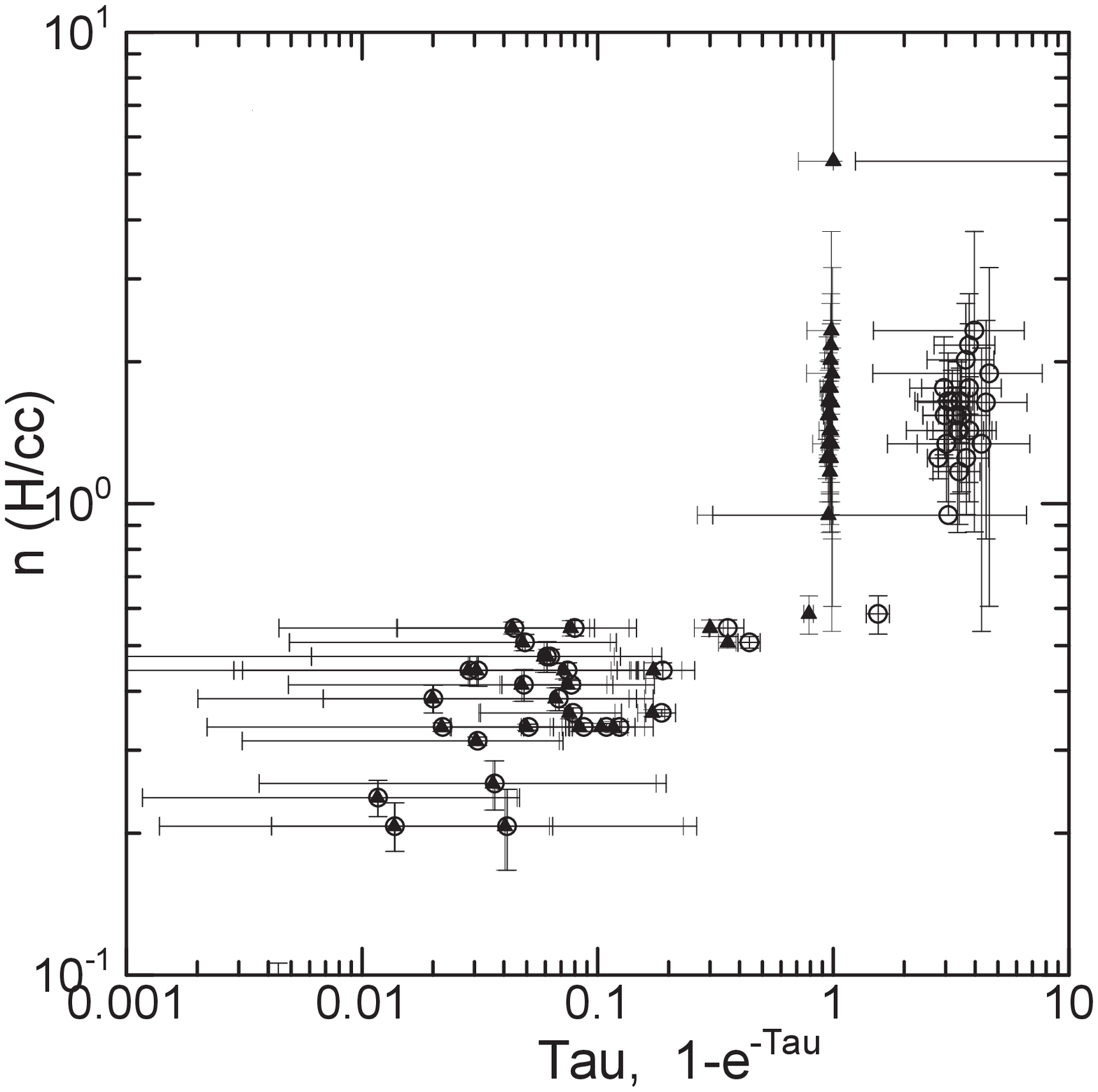}  
\end{center}
\caption{(Top) Plot of $\Ts$ and $n$ against $\tau$ (circles) and $1-e^{-\tau}$ (triangles). The straight line is an approximate fit to the present plot. The dashed and dotted lines are fits to Kuchar and Bania (1990) and Liszt et al. (1983), respectively. (Bottom) Plot of $n$ against $\tau$ (circles) and $1-e^{-\tau}$ (triangles)   } 
\label{mutual} 
	\end{figure}

Figure \ref{mutual} shows mutual relations among $\Ts$, $n$, and $\tau$. At small optical depths, $\Ts$ is a decreasing function of $\tau$, consistent with the results by Dickey et al. (1978) and Kuchar and Bania (1990). The straight line in the figure is an approximate fit to the present plot as 
\begin{equation}
\Ts  \sim 110 (1-e^{-\tau})^{-1} [{\rm K}]. 
\end{equation}
The dashed line in the figure is the result by Kuchar and Bania (1990). While the slope is almost identical, the present $\Ts$ is systematically higher. These may be further compared with the well established relation for HI clouds, $\Ts = 40 (1-e^{-\tau})^{-0.5}$, obtained by Liszt et al. (1983). The differences would be reasonable, if we consider the expected higher $\Ts$ in broadly distributed disk gas than that in individual HI clouds having higher density and lower kinetic temperature. 

The bottom panel of figure \ref{mutual} shows plots of $n$ against $\tau$. The density stays at $\sim 0.3 - 1$ H cm$^{-3}$ at $\tau <\sim 1$, but it increases to $\sim 1-2$ H cm$^{-3}$ at higher depth of $\tau \sim >1$. In high-$\tau$ conditions, the HI gas tends to be denser than that calculated for optically thin assumption.

	\begin{figure} 
\begin{center}       
\includegraphics[width=7cm]{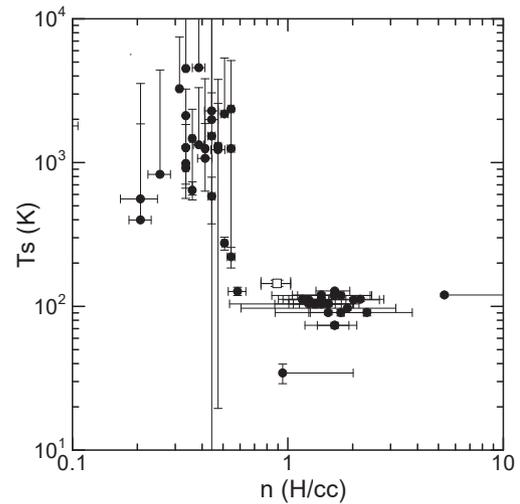} 
\end{center}
\caption{Plot of $\Ts$ against $n$.  }
\label{Tn} 
	\end{figure} 

Figure \ref{Tn} shows a plot of $\Ts$ against $n$. The warm and cold gases show a clear anti-correlation. The denser gas has lower spin temperature at $\Ts \sim 110$ K, and lower density gas has higher temperature at $\sim 2000$ K.

\begin{figure} 
\begin{center}      
\includegraphics[width=7cm]{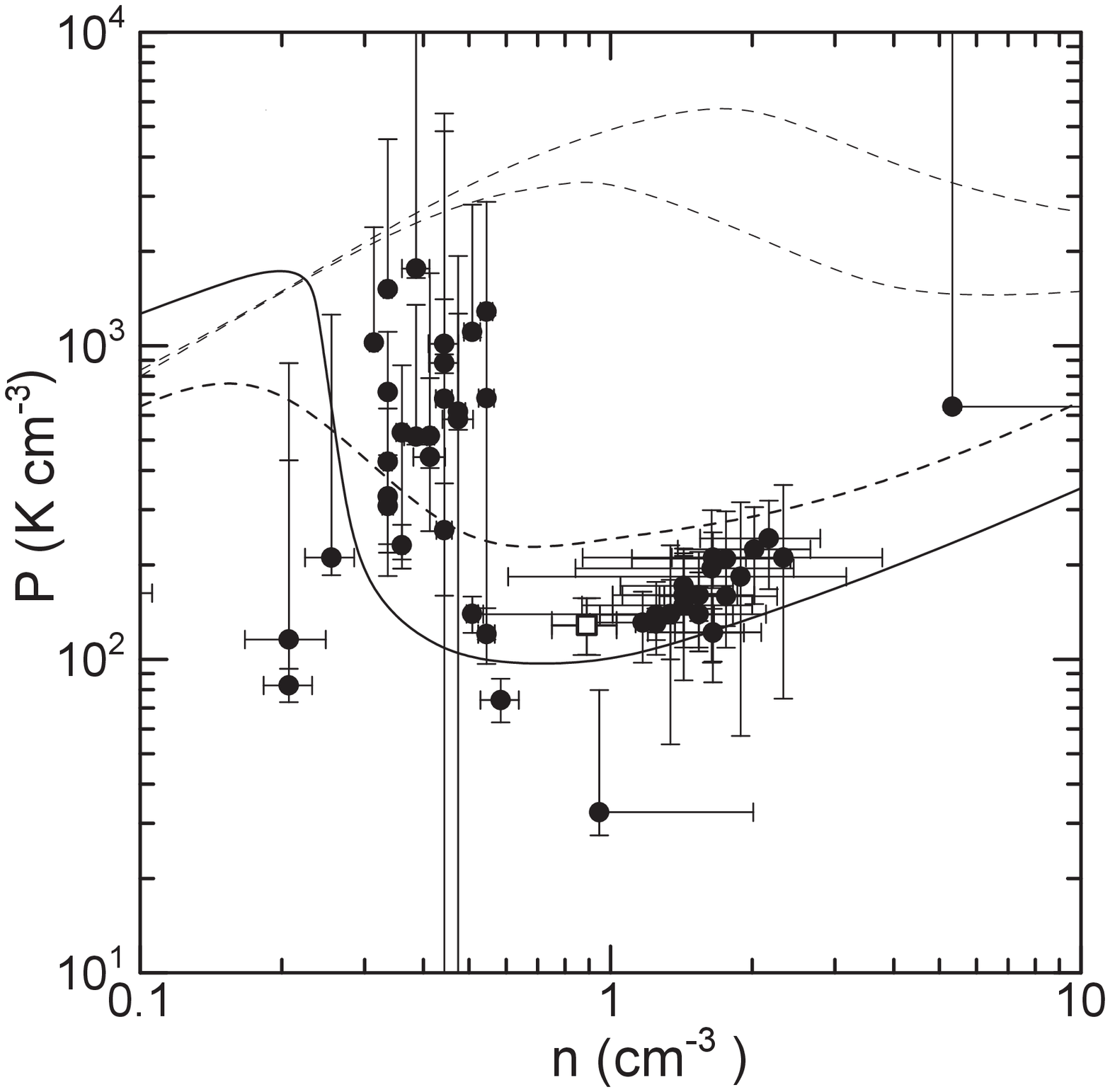} 
\includegraphics[width=7cm]{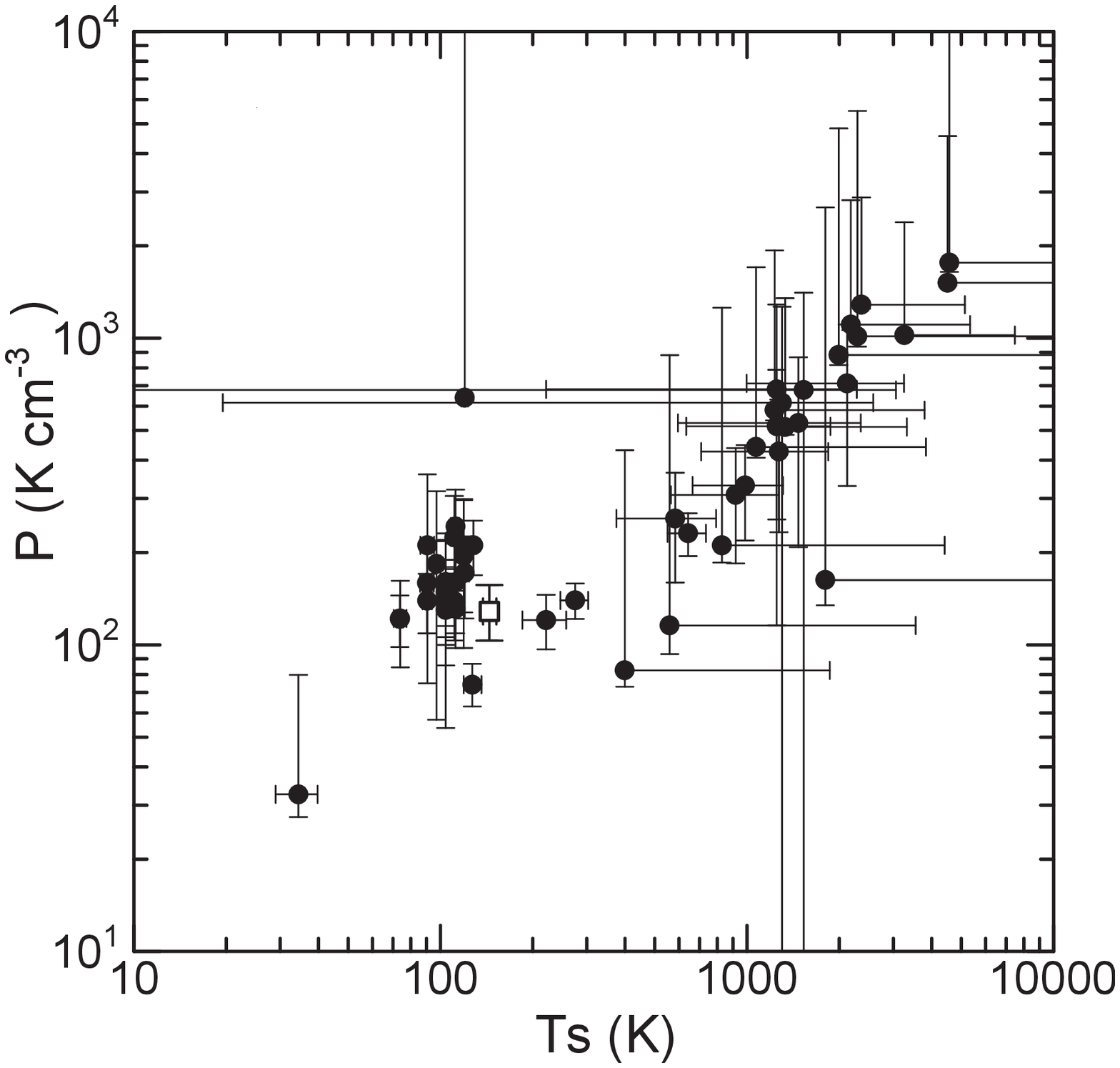}
\end{center}
\caption{[Top] Correlation between the pressure ($P=n\Ts$) and density $n$.  Solid and dashed lines in the $P-n$ relation shows the thermal equilibrium line obtained by Field et al. (1969) and Shaw et al. (2017), respectively. Upper two thin curves are calculations by Wolfire et al. (2003) at $R\sim 5$ kpc (dash) and 8.5 kpc (dot). [Bottom] Pressure plotted against $\Ts$.}
\label{pressure}
	\end{figure}

In figure \ref{pressure} we plot the pressure defined by $P=n\Ts$ against $n$ and $\Ts$. We insert a theoretical phase diagram for thermal equilibrium from Field et al. (1969), and a recent curve calculated for spin temperature by Shaw et al. (2017). The cold HI gas is tightly distributed in the high-density equilibrium region. On the other hand, the warm HI is distributed around unstable region with negative slope, $dP/dn<0$, which may imply that the presently detected warm HI in the inter-arm and GC is thermally unstable.

However, our plot is significantly displaced from the equilibrium curve calculated for kinetic temperature of pressure-balanced ISM in the interstellar radiation field (Wolfire et al. 2003) with an order of magnitude lower pressure.

\section{Discussion and Summary}

\subsection{HI masses: Revealing the hidden HI}
 
Figure \ref{fit_n} shows that the volume filling factors of the cold and warm HI gases are about the same, both $\sim 0.5$. Since the cold HI density is about 4 times the warm gas density, the total HI mass is about 2.5 times greater than the mass estimated by optically thin assumption. 

Assuming that the density $n$ approximately represents the azimutally averaged density at galacto-centric distance $R$, we may estimate the HI mass by integrating the density over the disk's volume,
\begin{equation}
M(R)=2 \pi   \mh \int_{-\infty}^{\infty} \int_0^R  n  r dr dz,
\label{mass}
\end{equation}
where $z$ is the vertical distance from the galactic plane. Using the full scale thickness of the HI disk (Nakanishi and Sofue 2016),
\begin{equation}
h(R)=110 \ {\rm exp}(+R/4.5 \ {\rm kpc}) \ [{\rm pc}],
\end{equation} 
we may rewrite equation (\ref{mass}) as
\begin{equation}
M=2 \pi \mh \int_0^{R_0}  n  h(r) r dr.
\end{equation} 

Replacing the integration by summation of the densities in figure \ref{fit_n} multiplied by tbe areas of rings, we obtain the cold, warm and total HI masses inside $R_0=8$ kpc, respectively, to be
$M_{\rm cold \ HI}\sim  1.47 \times 10^9 \Msun$, 
$M_{\rm warm \ HI} \sim 3.2 \times 10^8 \Msun$, and
$M_{\rm total \ HI} \sim 1.79 \times 10^9 \Msun$.
Here, we approximated the missing densities in the rings without meaningful fitting by the densities in the inner neighboring rings. 
The total HI mass is $\sim 2$ times that estimated by Nakanishi and Sofue (2016) ($\sim 0.9\times 10^9\Msun$) using the commonly used formula for optically thin HI,
\begin{equation}
N_{\rm HI}=\X \int \Tb dv.
\end{equation}
The HI mass may be also compared with the dynamical mass, $M_{\rm dyn}=R_0^2 V_0^2/G \sim 1.05\times 10^{11}\Msun$ with $V=238$ \kms and $R_0=8$ kpc, yielding an HI mass fraction of $\sim 0.017$.

\subsection{Comparison of the density profiles with other works}

	\begin{figure} 
\begin{center}      
\includegraphics[width=8cm]{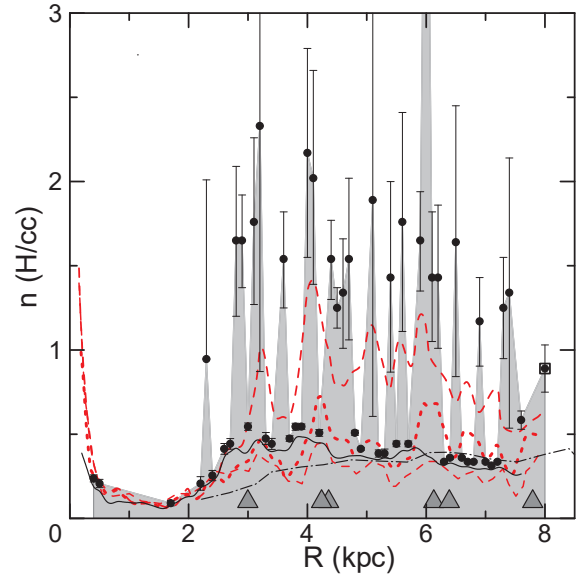}     
\end{center}
\caption{Density profiles as a function of $R=R_0\ |\sin \ l|$ (galacto-centric distance). [Grey area with circles] Present work; [Solid line] ibid for optically thin limit;  [Red dash from upper to lower] Marasco et al. (2017) for $\Ts=80$ K, 152 K, and 8000 K; [Dash-dot] Burton (1976). The distances are adjusted to $R_0=8$ kpc.  Big triangles mark the rings and spiral arms. }
\label{arms}
	\end{figure}

Although the here obtained quantities plotted against $R$ do not represent azimutally averaged values, we here try to compare them with the current studies of the radial distribution of the HI density. Figure \ref{arms} shows the $R$ dependence of volume density $n$ as compared with the current studies. The 3-kpc expanding gaseous ring, 4-kpc molecular ring and spiral arms No. 1, 3 and 4 from Nakanishi and Sofue (2006) are marked by big grey triangles, where No. 3 is the Sagittarius-Carina arm at $l\sim 50\deg$ ($R\sim 6.2$ kpc), and No.4  is the Scutum-Crux arm at $l\sim 307\deg$ ($R\sim 6.5$ kpc).

Our density profile for the optically thin limit generally agrees with the plots by Burton (1976) and the optically thin case assuming $\Ts=8000$ K by Marasco et al. (2017). These agreements confirm that the presentely derived HI density along the tangent-point circle may reasonably approximate the azimuthally averaged density. We may thus consider that the $R$ variations of the derived quantities approximately represent the true (azimuthally averaged) radial variations in the Galactic disk.

The density profile by Marasco et al. (2017) for optically thick case with $\Ts=80$ K (red dots) shows a similar variation to our cold HI profile, although their profile is much milder. Their 'fiducial' density profile for $\Ts=152$ K lies closer to the optically thin case. This may be explained by the assumed higher $\Ts$, which presumes more optically thin condition.

\subsection{Relation with the spiral arms}

\begin{figure} 
\begin{center}      
\includegraphics[width=7cm]{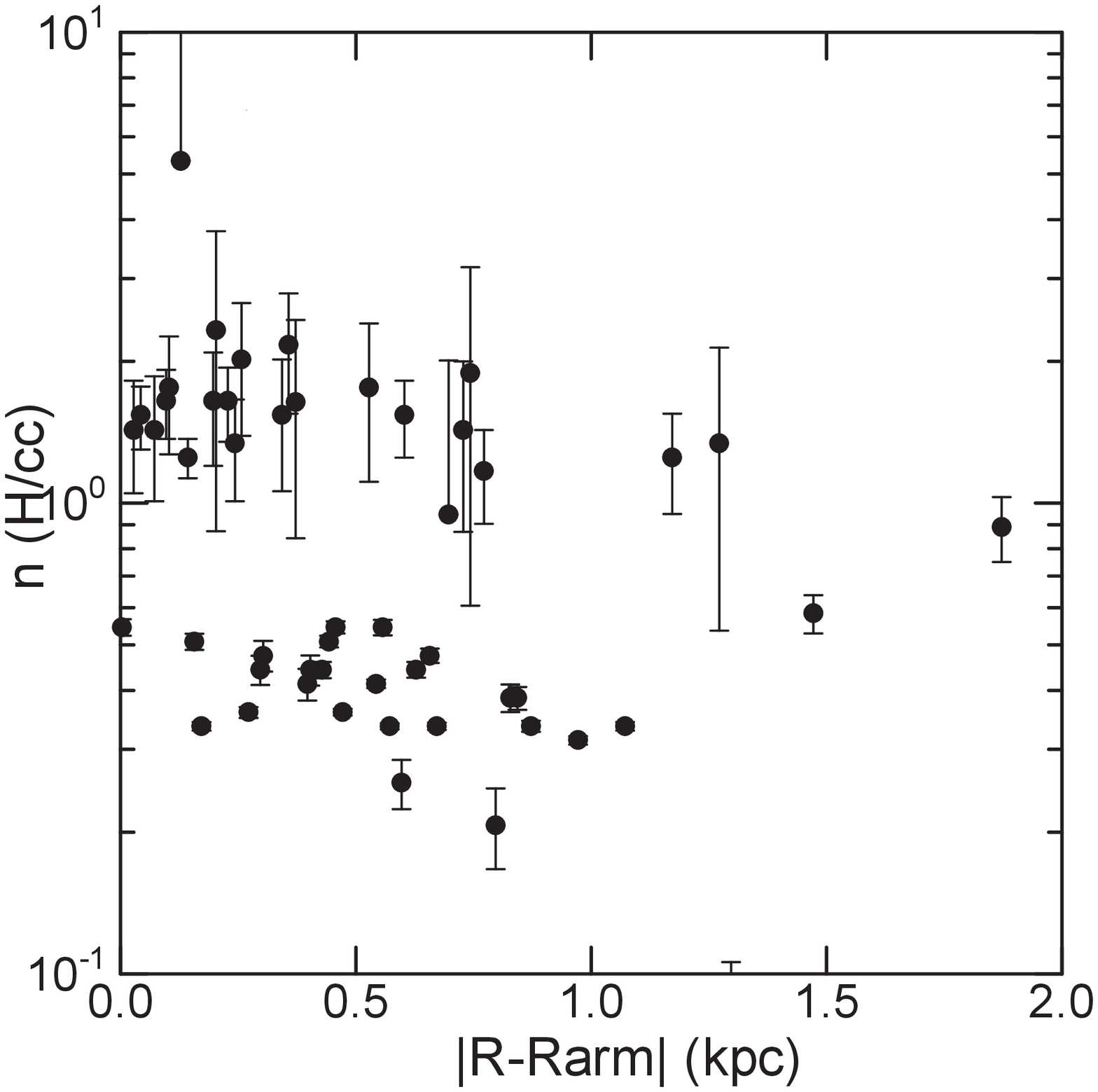}  
\includegraphics[width=7cm]{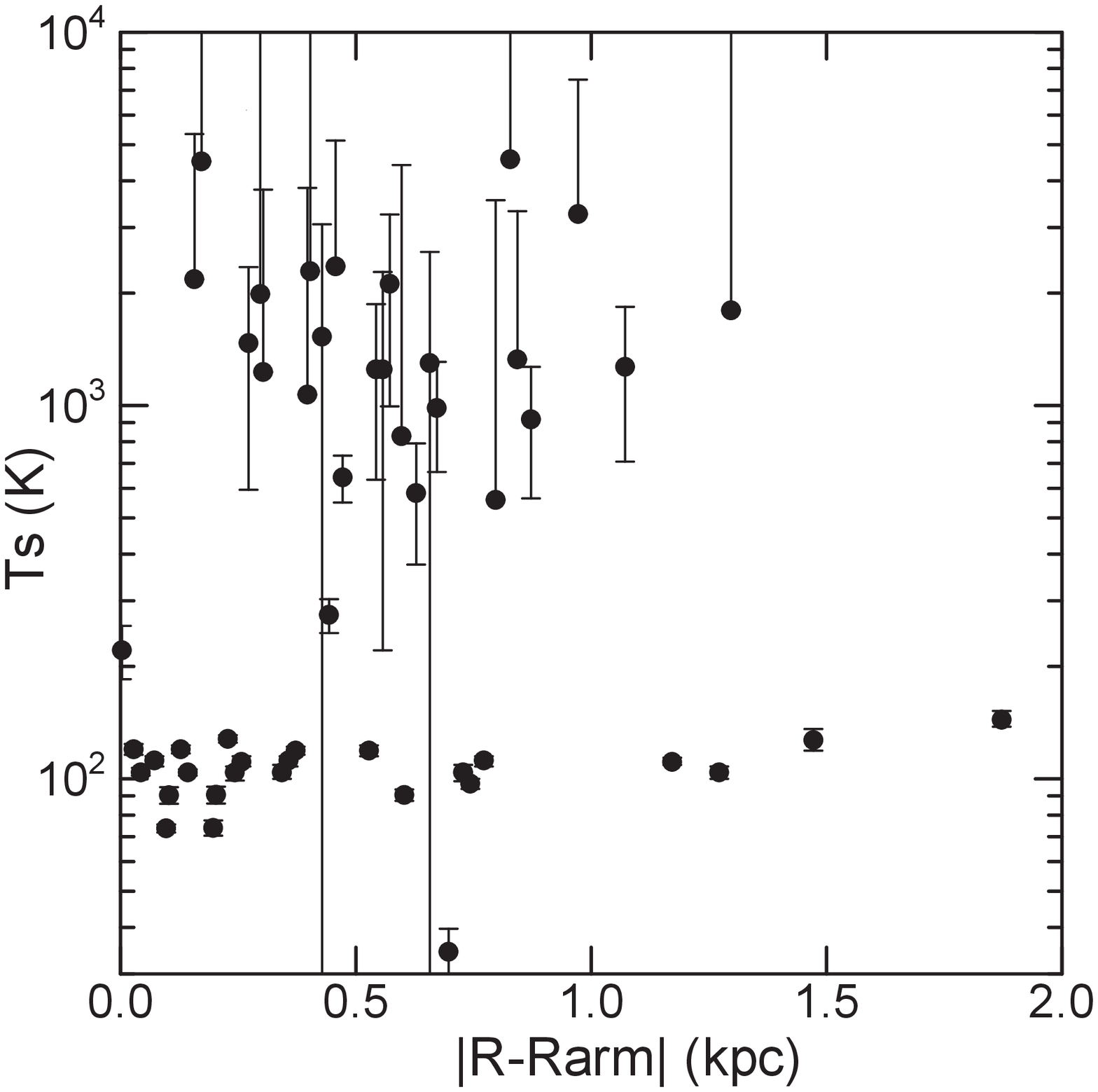}          
\end{center}
\caption{Density (top) and spin temperature (bottom) plotted against radial displacement, $\D$, of the measured ring from the nearest spiral arm as indicated by triangles in figure \ref{arms}. }
\label{darms}
	\end{figure}

There seems to exist a correlation between the best-fit density and the tangential directions of spiral arms as marked by the big triangles in figure \ref{arms}. In order to examine the correlation, we plot $n$ and $\Ts$ against radial displacement, $\D$, of the analyzed rings from the nearest arms in figure \ref{darms}. The figure shows that the warm and cold HI densities decrease with $\D$ at $e$-folding length of about 1 kpc. The spin temperatre $\Ts$ of cold HI is nearly constant, while slightly increases with the distance from the arms. The spin temperature of warm HI seems also to be increasing with the distance, but the scatter and errors are too large for measuring the slope.

\begin{figure} 
\begin{center}       
\includegraphics[width=7cm]{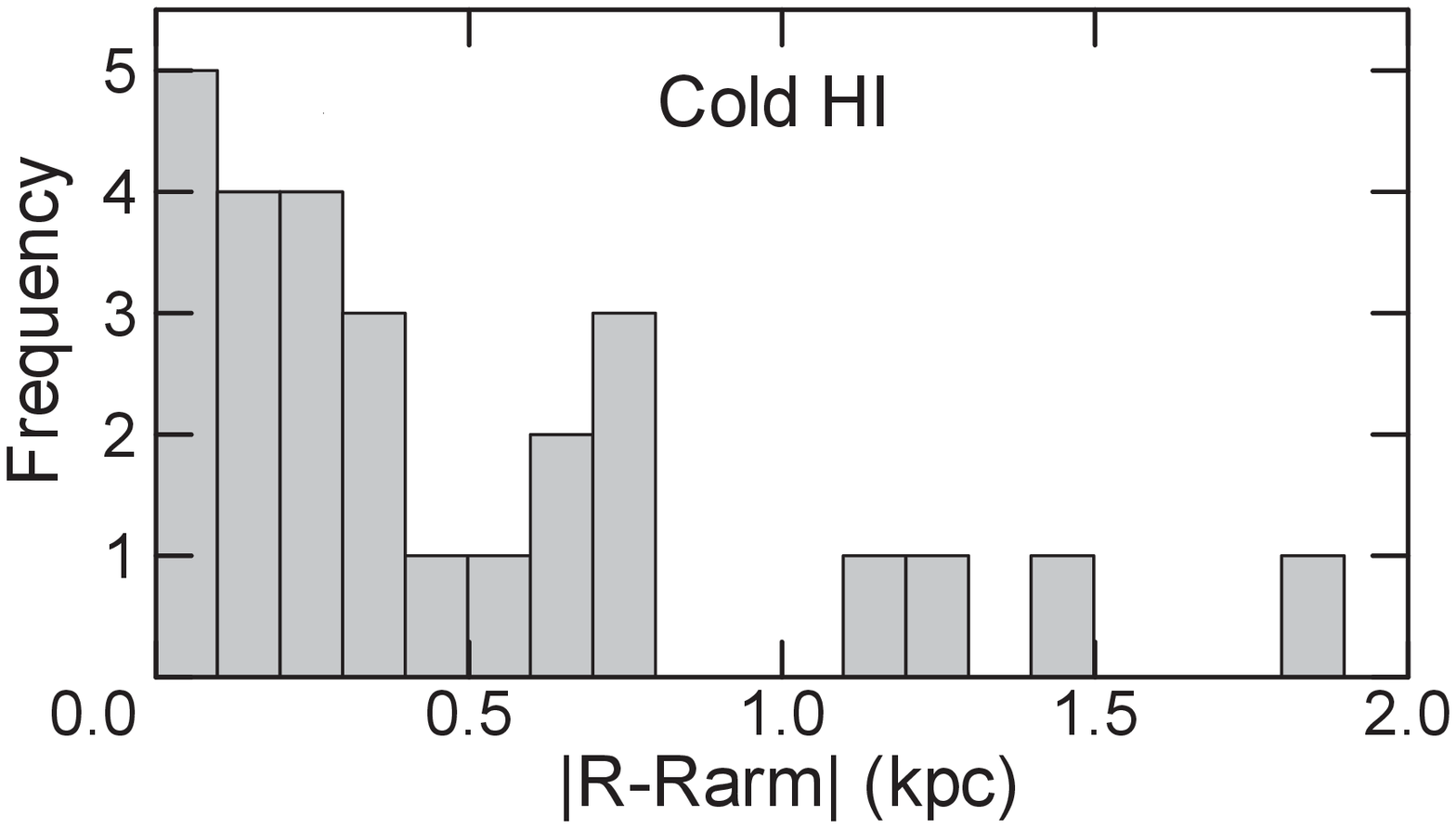} 
\includegraphics[width=7cm]{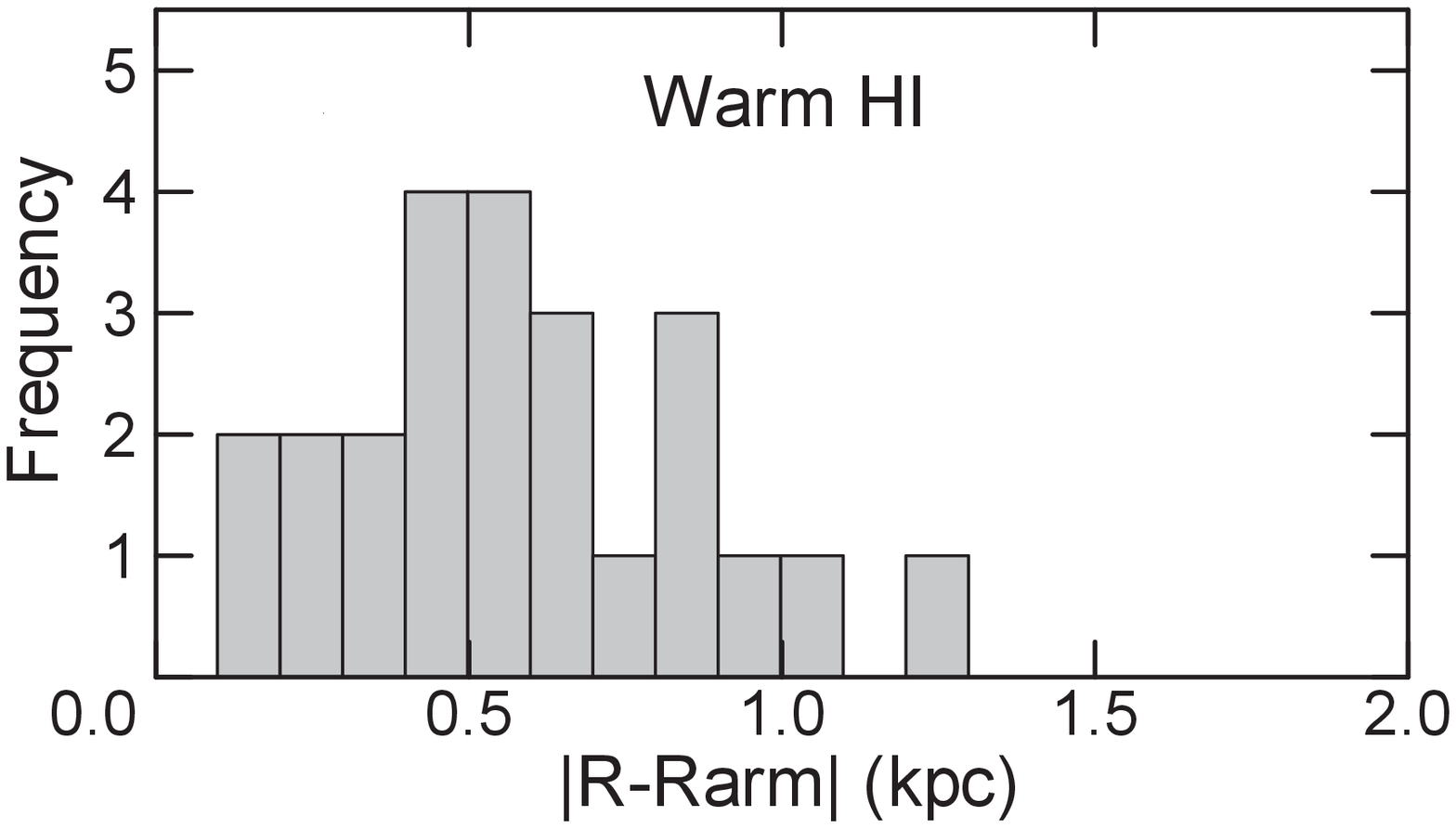}        
\end{center}
\caption{Histograms of the frequencies of $\D$ of the cold and warm HI.}
\label{darmhisto}
	\end{figure}

Figure \ref{darmhisto} shows histograms of the frequencies of appearence of cold and warm HI as a function of $\D$. The figure shows that cold HI gas is concentrated toward the arms at $\D \le \sim 0.5$ kpc, while warm HI is distributed farther out of the arms peaking around $\D \sim 0.5$ kpc.

\begin{figure} 
\begin{center}       
\includegraphics[width=6cm]{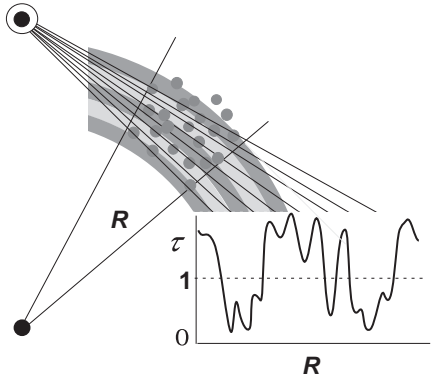}         
\end{center}
\caption{Illustration to explain how the optical depth varies across the arms with low $\Ts$ and high $n$ (cold HI; grey area and clouds), and interarm regions with high $\Ts$ and low $n$ (warm HI).}
\label{armCandW}
	\end{figure}

        Besides the global correlation, figure \ref{arms} shows shorter scale rapid variation with a typical wavelength of $0.1-0.2$ kpc. This may be due to clumpy HI gas distribution around the arms. Figure \ref{armCandW} illustrates possible distribution of HI gas and clouds around tangent points, where the lines of sight encounter a number of clouds inside the arms, but few or noe in the inter arm. The clumpiness causes short-scale variation of the optical depth, and hence the fitted $n$ and $\Ts$ varies accordingly.

\subsection{Coupling of the density and velocity dispersion} 

We used equations (\ref{eqTb}) and (\ref{eqtau}), where the latter reduces to equation (\ref{eqtaudvdr}), to solve for the parameters $\Ts$ and $n$. We here recall that $n$ and $\sigv'$ appear always in the form of $n/\sqrt{\sigv'}$, so that we can rewrite the equation using only two parameters, $\Ts$ and $n'=n/\sqrt{\sigv'}$. 
Since the term including $\dlnv$ is an order of magnitude smaller than $\sigv$,  the result will approximately coincide with that so far displayed by replacing $n$-axes in the displayed plots by $n'$ and multiplying $1/\sqrt{10}$, but in unit of \Hcc (\kms)$^{-1/2}$.  
        
We may further try, though not here, to correct the presently obtained HI density for observed $\sigv$ by multiplying $(\sigv/10 {\rm km\ s^{-1})^{-1/2}}$. The results would be changed only slightly by a factor of $\sim 0.9$ to $\sim 1.1$ according to $\sigv \sim 8$ \kms near the Sun to $\sim 12$ \kms in the inner region as measured by Marasco et al. (2017).
 
\subsection{Advantage of the method}
In this paper we could determine the density $n$ and spin temperature $\Ts$ of HI at the same time. The spatial position of measured HI gas is precisely known at a tangent point of galactic rotation. 
The large number and continuous sampling of input data would be also an advantage for studying the broadly distributed galactic HI gas. 

The number of independent data points were as large as $N\sim (\delta l/\theta_{\rm B}) N_{\rm Q} N_{\rm P}$, where $\delta l$ is the longitude interval corresponding to $\Rwidth=0.1$ kpc, $\theta_{\rm B}$ is the beam width, $N_{\rm Q}=2$ is the number of observed quadrants of the galactic disk, and $N_{P}=2$ is the number of observations (LAB and GASS). The number in each radial bin for the least-square fitting was greater than 12 (degree of freedom 10), sufficiently large for the meaningful fitting. 

\subsection{Limitation and uncertainty}

In contrast to the cold HI, the warm HI showed large errors and scatter. This may be due to contamination of the neighboring cold gas, or by its intrinsic fluctuation. This could be clarified by applying the present method to higher resolution data observed with 100-m class telescopes and/or interferometers.

We represented the background continuum emission by equation (\ref{TcBack}), assuming that the radio disk is axisymmetric around the GC. This is obviously too much simplified. It would be necessary to analyze the east and west quadrants of the disk individually, if we want to discuss the spiral arms in further detail. 
Also ignored here is the contribution from individual radio sources, which may be located either in front or beyond the TP, while we here divided them into two equal emissions. All these problems may be rather easily solved using higher-resolution LV data and radio source catalogs, and are subject for the future.

It should be noted that these inconvenient treatments of the continuum background may be another cause of the unstable (lacking) fitting results in several directions in figures \ref{fit_Ts} to \ref{fittau}.

\subsection{Convergence of the method}

It would be trivial that the best-fit (or most probable) values of $\Ts$ and $n$ approximately satisfy equation \ref{eqTb} by minimizing $\chi^2$ through equation \ref{eqchisq}. However, in order to doubly check the reliability of the method, it may be worth to check if it works to reproduce an artificially given model. 

\begin{figure} 
\begin{center}       
(A) Warm HI ~~~~~~~~~~~~~~~~~~~~~~~~~~~~~ (B) Cold HI 
\\
\hskip-5mm\includegraphics[width=43mm]{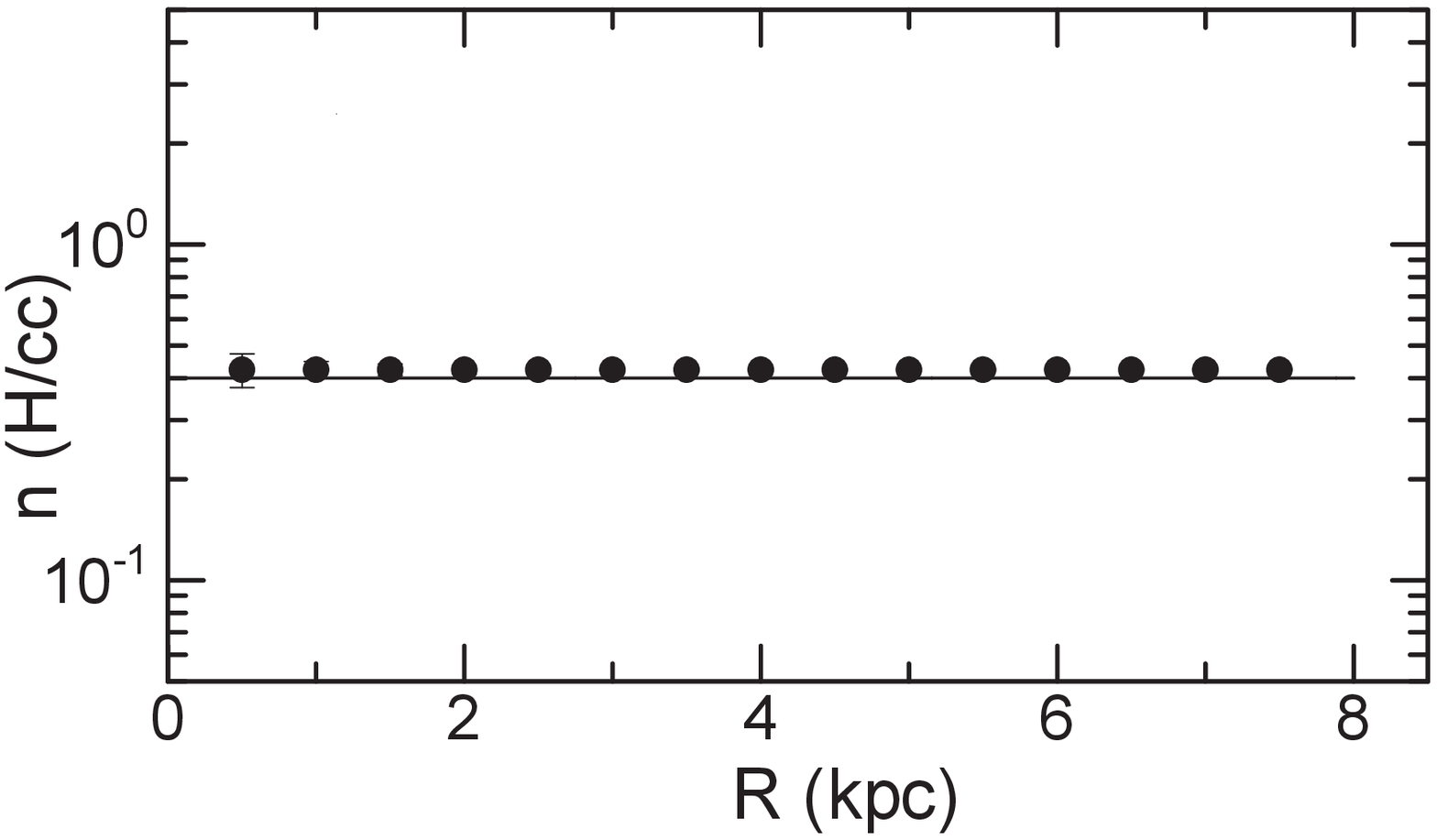} 
\includegraphics[width=43mm]{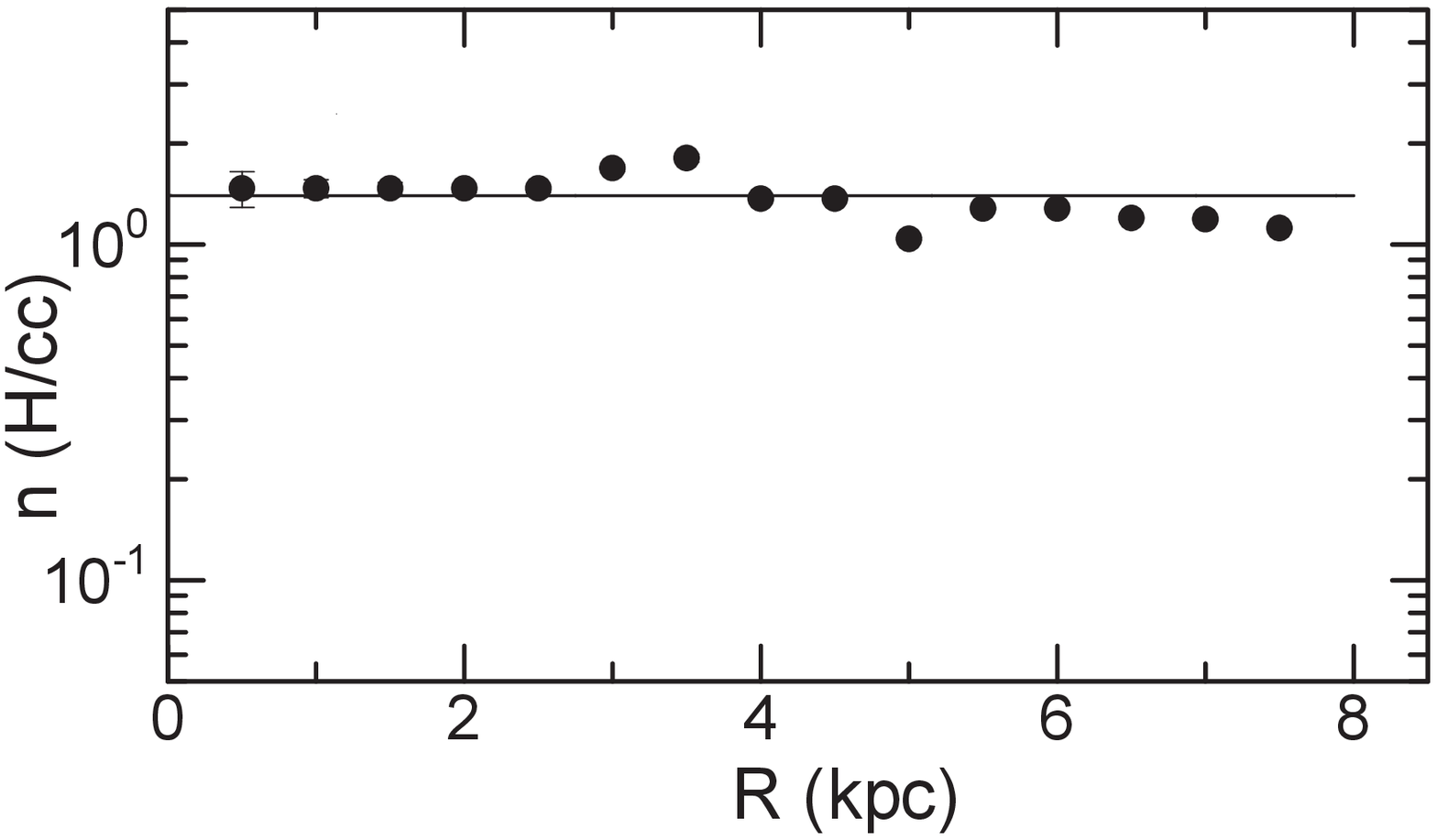} \\     
\hskip-5mm\includegraphics[width=43mm]{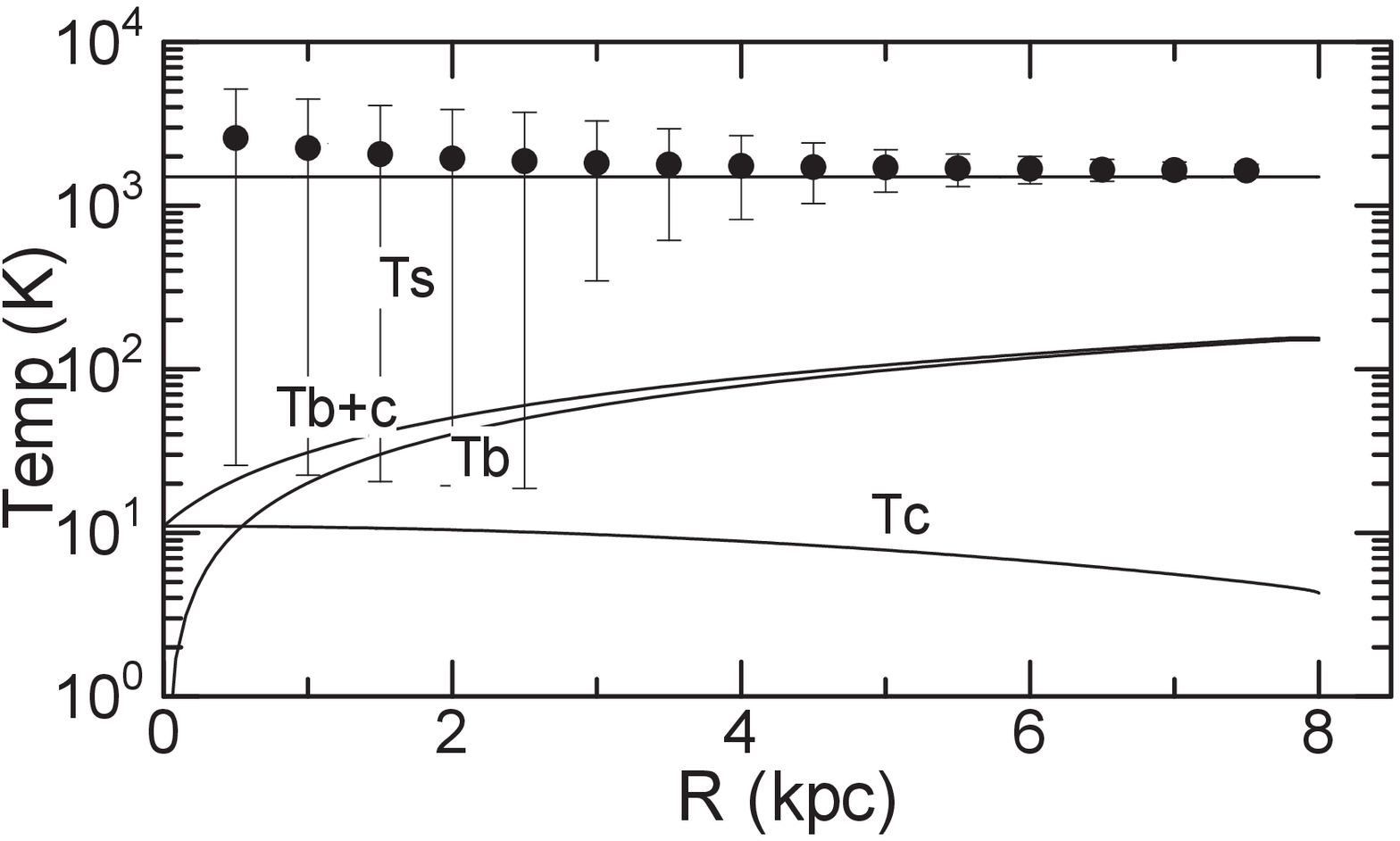}  
\includegraphics[width=43mm]{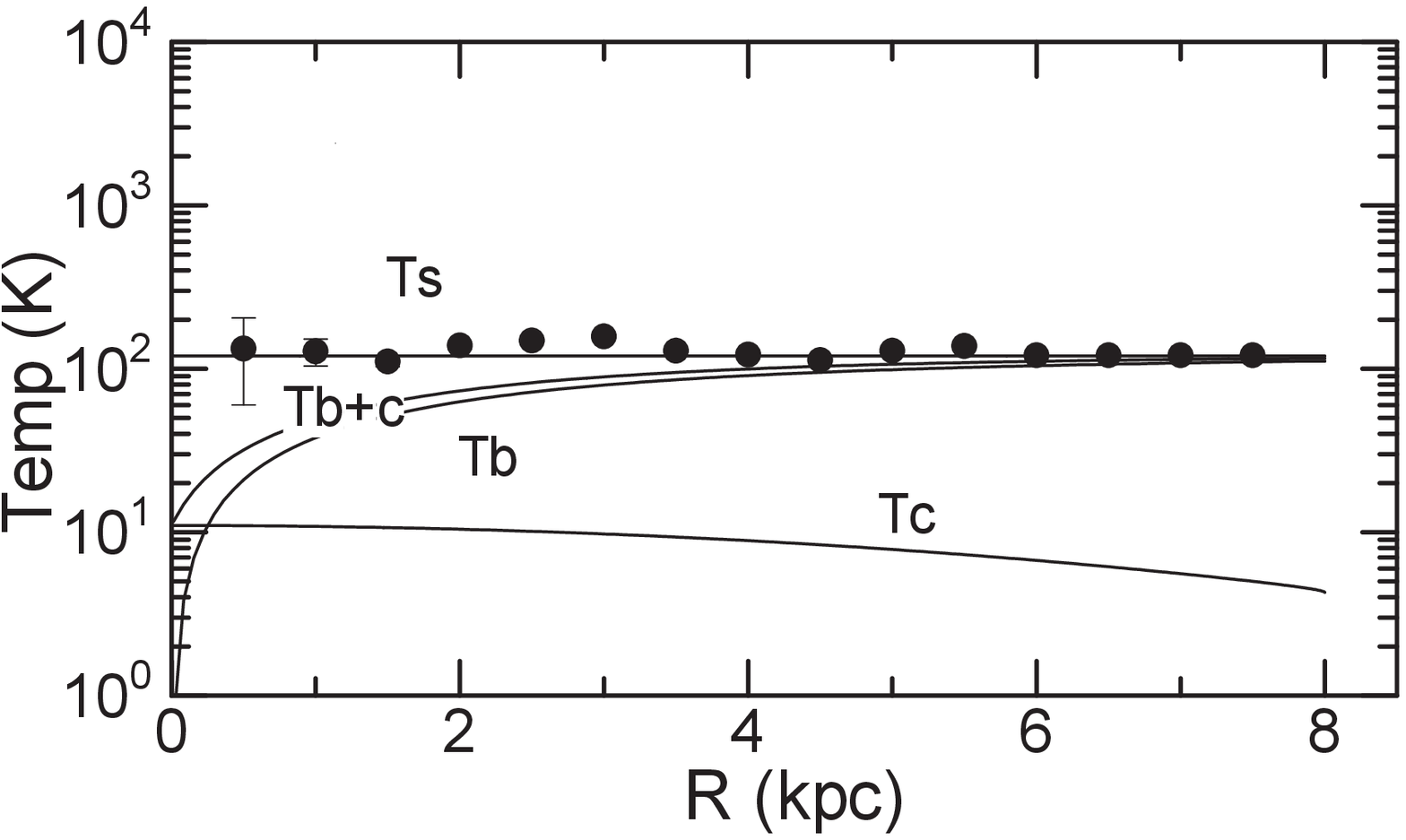}  \\ 
\hskip-5mm\includegraphics[width=43mm]{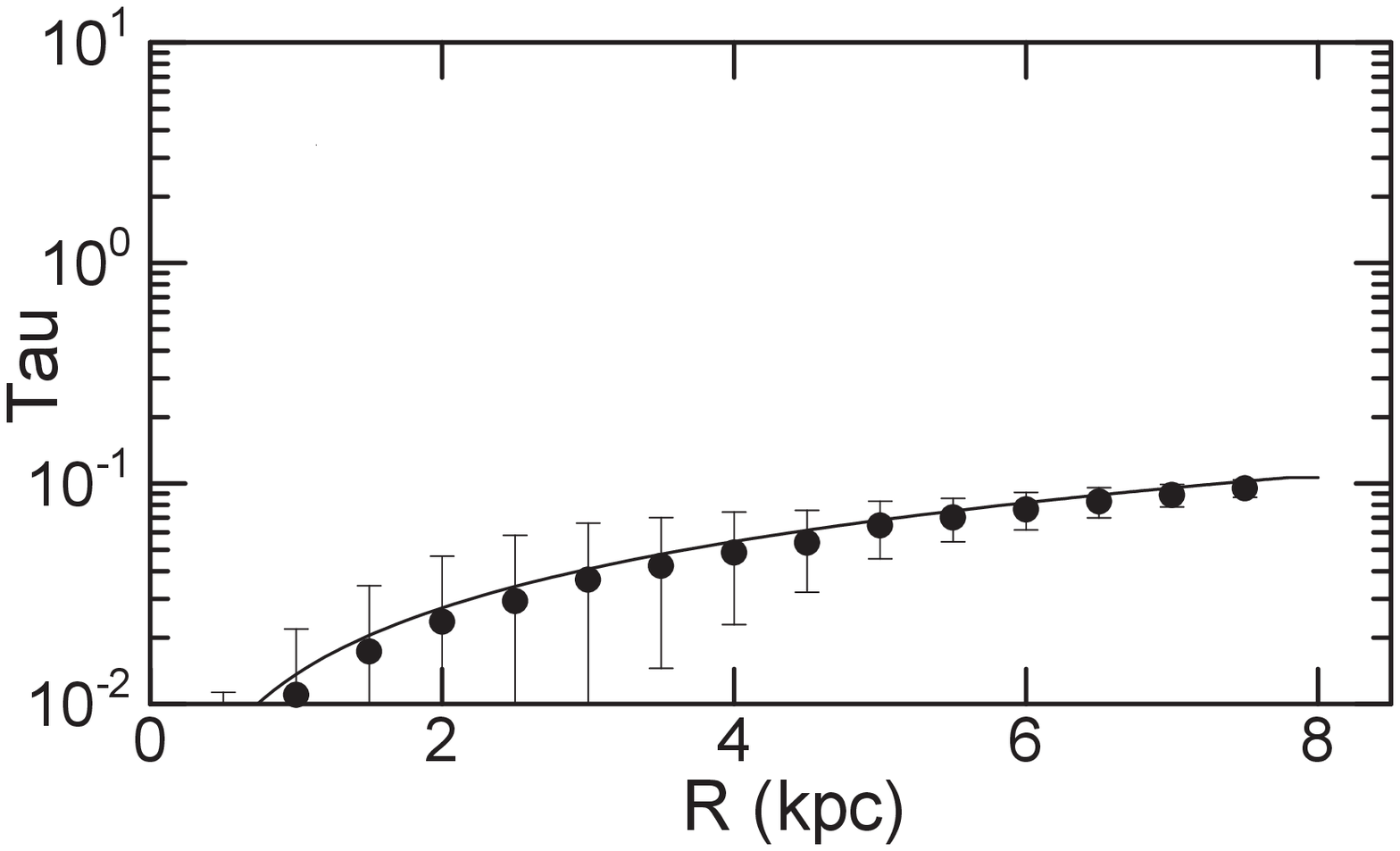} 
\includegraphics[width=43mm]{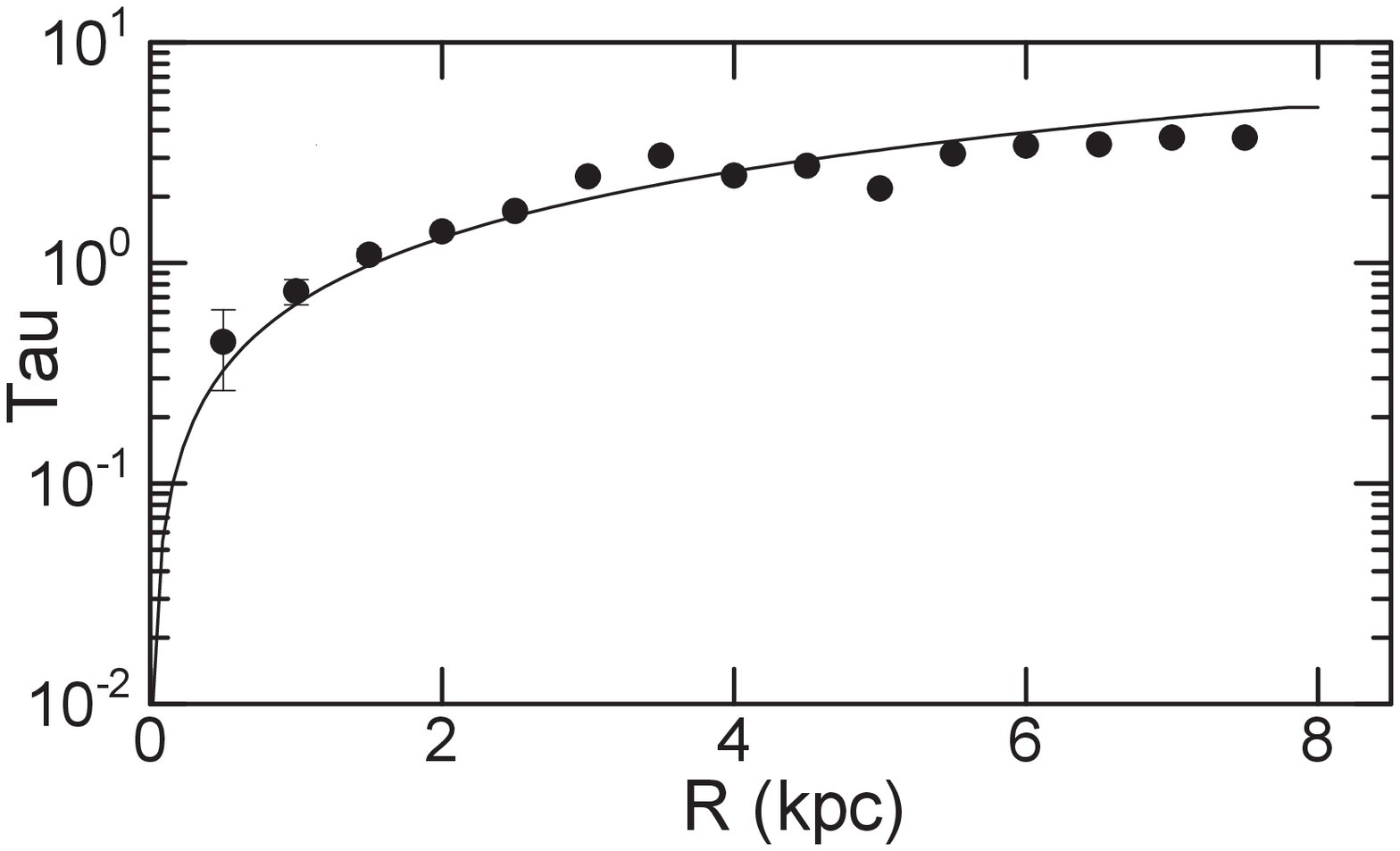}  
\end{center}
\caption{Convergence of the least-$\chi^2$ results (black dots) for one-temperature HI models.  (A) $\Ts=1500$ K, $n=0.4$ \Hcc, and flat RC at 220 \kms. (B) Same but $\Ts=110$ K and $n=1.4$ \Hcc. 
(Top) Density plotted against $R$.
(Middle) Temperatures. 
(Bottom) Optical depth. }
\label{model}
	\end{figure}

As templates we consider two models: (A) One-temperature warm HI disk with $\Ts=1500$ K and constant density of $n=0.4$ \Hcc; and (B) cold HI disk with $\Ts=110$ K and $n=1.4$ \Hcc. We assume a constant velocity dispersion of $\sigv=10$ \kms and flat rotation curve with $V=220$ \kms. The solid lines in the middle panels of figure \ref{model} show assumed $\Ts$ and calculated $\Tb$ for the flat density model. The 1.4 GHz continuum emission $\Tc$ was modeled by a Gaussian function around the GC plus constant 3 K emission. 

Using thus calculated model $\Tb$ and $\Tc$ as the input observables, we performed the least-$\chi^2$ fitting to derive the best-fit $n$ and $\Ts$ at various rings with $\Rwidth= 0.1$ kpc. Figure \ref{model} plots the fitting results compared with the input curves, where the method reasonably reproduces the given model profiles. However, results in the central region show large errors and scatter, which is due to the small optical depth mainly caused by increasing $V/R$ in equation (\ref{eqtau}) toward the GC, This means that the method is less reliable in regions having low density and/or low optical depth. Such tendency is readily seen in figures \ref{fit_n} and \ref{fit_Ts} as missing plots in the low HI density region around the GC.

\subsection{Summary}

Extending the VDR-$\chi^2$ method, we developed a method to determine the spin temperature and volume density simultaneously at tangent points of galactic rotation in the Galactic disk. The averaged values of $\Ts$ and $n$ in the galactic disk from $R=3$ to 8 kpc were obtained to be $\Ts=106.7 \pm 16.0$ K and $n=1.53\pm 0.86$ \Hcc for cold HI (CNM), and $1721 \pm 1060$ K and $0.38 \pm 0.10$ \Hcc for warm HI (WNM). 
 
The variation of $\Ts$ and $n$ against $R$  showed that the spin temperature of cold HI stays nearly constant at $\sim 110$ K $R=3$ to 7.5 kpc, whereas $\Ts$ of warm HI $\Ts$ is more scattered with large error. The GC region and the HI hole inside $\sim 2$ kpc are dominated by warm HI.

 Detailed $R$-distributions of $n$ and $\Ts$ showed that the cold, dense HI is located in the spiral arms and rings, whereas the warm, low-density HI is found in the inter-arm regions. Figure \ref{arms} shows the density profile in order to compare the density peaks and valleys with the arm positions. 
 
The cold HI gas is in average $\sim 4$ times denser than the warm HI, which means that the galactic HI mass is significantly larger than the currently estimated value on the optically thin assumption.   It is also found that the density contrast between arms and inter-arm region is as high as $\sim 4$ .
 The total HI mass inside the solar circle is estimated to be 2.5 times greater than that calculated for optically thin assumption.

The cold and warm HI gases are distributed along the thermal equilibrium locus in the phase diagram in the spin temperature-pressure plane. The cold HI gas is distributed in thermally stable region, whereas warm HI is in the unstable region.

\section*{Acknowledgments}
The author expresses his sincere thanks to the authors of the LAB (Dr. Kalberla et al.) and GASS (Dr. \Mc et al.) HI line surveys, and the SVE radio continuum survey (Dr. P. and W. Reich, et al) for the archival FITS formatted data cubes.

\end{document}